\documentclass[galaxies,review,accept,moreauthors,pdftex]{Definitions/mdpi} 
\usepackage[normalem]{ulem}

\newcommand\aap{A\&A}                
\newcommand\aj{AJ}                   
\newcommand\apj{ApJ}                 
\newcommand\apjl{ApJ}                
\newcommand\apjs{ApJS}               
\newcommand\araa{ARA\&A}             
\newcommand\mnras{MNRAS}             
\newcommand\nat{Nature}              
\newcommand\prd{Phys. Rev.~D}        
\newcommand\prl{Phys. Rev.~Lett.}    
\newcommand\pasj{PASJ}               
\newcommand\physrep{Phys.~Rep.}      
\newcommand\sovast{Soviet~Ast.}      
\newcommand\ssr{Space Sci. Rev.}     
\firstpage{1} 
\pubvolume{1}
\issuenum{1}
\articlenumber{0}
\pubyear{2021}
\copyrightyear{2020}
\datereceived{} 
\dateaccepted{} 
\datepublished{} 
\hreflink{https://doi.org/} 

\Title{Gamma-ray and Neutrino Signals from Accretion Disk Coronae of Active Galactic Nuclei}

\TitleCitation{Title}

\Author{Yoshiyuki Inoue $^{1,2,3,*}$\orcidA{}, Dmitry Khangulyan $^{4}$\orcidB{} and Akihiro Doi $^{5,6}$\orcidC{}}

\AuthorNames{Yoshiyuki Inoue, Dmitry Khangulyan and Akihiro Doi}

\AuthorCitation{Inoue, Y.; Khangulyan, D.; Doi, A.}

\address{%
$^{1}$ \quad Department of Earth and Space Science, Graduate School of Science, Osaka University, Toyonaka, Osaka 560-0043, Japan;\\
$^{2}$ \quad Interdisciplinary Theoretical \& Mathematical Science Program (iTHEMS), RIKEN, 2-1 Hirosawa, Saitama 351-0198, Japan;\\
$^{3}$ \quad Kavli Institute for the Physics and Mathematics of the Universe (WPI), The University of Tokyo, Kashiwa 277-8583, Japan;\\
$^{4}$ \quad Department of Physics, Rikkyo University, Nishi-Ikebukuro 3-34-1, Toshima-ku, Tokyo 171-8501, Japan;\\
$^{5}$ \quad Institute of Space and Astronautical Science JAXA, 3-1-1 Yoshinodai, Chuo-ku, Sagamihara, Kanagawa 252-5210, Japan;\\
$^{6}$ \quad Department of Space and Astronautical Science, The Graduate University for Advanced Studies (SOKENDAI),3-1-1 Yoshinodai, Chuou-ku, Sagamihara, Kanagawa 252-5210, Japan}

\corres{Correspondence: yinoue@astro-osaka.jp}

\abstract{To explain X-ray spectra of active galactic nuclei (AGN), non-thermal activity in AGN coronae such as pair cascade models has been extensively discussed in the past literature. Although X-ray and gamma-ray observations in the 1990s disfavored such pair cascade models, recent millimeter-wave observations of nearby Seyferts establish the existence of weak non-thermal coronal activity. Besides, the IceCube collaboration reported NGC 1068, a nearby Seyfert, as the hottest spot in their 10-yr survey. These pieces of evidence are enough to investigate the non-thermal perspective of AGN coronae in depth again. This article summarizes our current observational understandings of AGN coronae and describes how AGN coronae generate high-energy particles. We also provide ways to test the AGN corona model with radio, X-ray, MeV gamma-ray, and high-energy neutrino observations.}


\begin{document}
\section{Introduction}

Back in 2013, the IceCube collaboration reported the evidence for astrophysical neutrinos with energies in the 30~TeV-1~PeV range \citep{IceCube2013PhRvL.111b1103A}. The detected neutrinos show isotropic distribution with the flux level of $\approx10^{-8}~{\rm GeV\ cm^{-2}\ s^{-1}\ sr^{-1}}$ \citep{IceCube2014PhRvL.113j1101A}. This discovery indicates that TeV-PeV neutrinos travel to the Earth from the distant Universe. Although various theoretical models have attempted to explain the cosmic TeV-PeV neutrino fluxes even before the operation of IceCube \citep[see, e.g.,][for reviews]{Murase2017nacs.book...15M, Ahlers2018PrPNP.102...73A}, the lack of clear source identification hampers our understandings of the origin of the TeV-PeV neutrinos. 

In 2018, the IceCube collaboration, together with the other electromagnetic wave observatories, reported a possible spatial and temporal coincidence of a neutrino event with a blazar flare TXS~0506+056 \citep{IceCube2018Sci...361..147I, IceCube2018Sci...361.1378I}. This observation provides the first evidence of a blazar as a cosmic neutrino factory. However, stacking analysis of blazars show that they can afford only up to $\sim30$\% of the measured cosmic TeV-PeV neutrino background flux \citep{IceCube2017ApJ...835...45A}. Therefore, even after the first possible source identification, the origin of the cosmic neutrino background flux was still veiled in mystery.

Very recently, the accumulation of 10-yr IceCube survey data revealed the existence of a neutrino hot spot toward the direction of NGC~1068 with a $2.9\sigma$ confidence level \citep{IceCube2020PhRvL.124e1103A}. NGC~1068 is one of the nearest \citep[$\sim14$~Mpc;][]{Tully1988} and the best-studied Seyfert~2 galaxies in broadband \cite{Bauer2015, Marinucci2016,Pasetto2019}. A Seyfert~2 galaxy is a type of active galactic nucleus (AGN) population, where the central engine is supposed to be blocked by the dusty torus. The surface density of Seyferts is $\approx {\rm 2\times10^4\ deg^{-2}}$ \citep{Luo2017ApJS..228....2L}, which is about four orders of magnitude higher than that of blazars \citep{Marcotulli2020ApJ...896....6M}. This possible neutrino signal from NGC~1068 may indicate that Seyferts dominate the cosmic neutrino sky. An immediate question is how Seyferts generate neutrinos. 

Connecting gamma-ray and neutrino measurements would allow us to understand the neutrino production mechanism. Indeed, gamma-ray observatories such as {\it Fermi} and {MAGIC} have detected GeV-TeV gamma-ray photons from NGC~1068 \citep{Fermi2012ApJ...755..164A, MAGIC2019ApJ...883..135A}. However, it turns out that the neutrino flux is brighter than the gamma-ray flux, which simple hadronuclear ($pp$) or photomeson ($p\gamma$) interpretation processes can not accommodate. 

Several theoretical models have already been proposed to explain the neutrino excess seen in NGC~1068, such as accretion disk corona \citep{Inoue2020ApJ...891L..33I, Murase2020PhRvL.125a1101M,Gutierrez2021arXiv210211921G, Anchordoqui2021arXiv210212409A}, the interaction of broad-line-region clouds with accretion disk \citep{Muller2020A&A...636A..92M}, and galactic cosmic-ray halo \citep{Recchia2021arXiv210105016R}. However, as always, for theoretical models, model uncertainties in all of these models are still non-negligible.

Here, non-thermal activity in AGN coronae has been studied in literature for several decades \citep[e.g.,][]{Zdziarski1986, Kazanas1986, Sikora1987,Begelman1990,Stecker1991PhRvL..66.2697S,Coppi1992MNRAS.258..657C,Ghisellini2004,Mastichiadis2005A&A...433..765M,Belmont2008A&A...491..617B,Poutanen2009ApJ...690L..97P,Kalashev2015,Inoue2019ApJ...880...40I,Murase2020PhRvL.125a1101M}. The predicted diffuse neutrino flux by one of the pioneering works was at the level of $\sim10^{-7}~{\rm GeV\ cm^{-2}\ s^{-1}\ sr^{-1}}$ \citep{Stecker1992}. Although this is significantly higher than the measured values, we note that it was more than 20 years before the first measurement. When those models are proposed, the lack of various observational information led to such a huge model uncertainty.  These results highlight that a better understanding of AGN coronae's properties is critical for understanding the neutrino production in AGN coronae.

The recent development of radio and X-ray telescopes now allows us to determine various physical parameters of AGN coronae, such as electron density, temperature, size, and magnetic field strength \citep{Fabian2015, Inoue2018}. ALMA observations also show evidence of weak non-thermal coronal activity, which establishes the existence of high energy particles in AGN coronae \citep{Inoue2020ApJ...891L..33I}. 

In this {\it Review}, we aim to summarize our current understanding of AGN coronae and describe how high energy signals arise from AGN coronae following \citet{Inoue2019ApJ...880...40I}. In \S.~\ref{sec:history}, we briefly look back at the research history of the high energy aspect of AGN coronae and summarize our current observational understandings of AGN coronae. In \S.~\ref{sec:model}, we describe the generation process of high energy particles in AGN coronae. In \S.~\ref{sec:g_nu_AGN}, we show the expected gamma-ray and neutrino signals and discuss the corona scenario interpretation of the neutrino signals from NGC~1068. In \S.~\ref{sec:CGNB}, the expected cosmic gamma-ray and neutrino background fluxes are described. In \S.~\ref{sec:discussion}, we compare available models and discuss the ways to test the coronal hypothesis. In \S.~\ref{sec:summary}, we summarize this review.

\section{Research History of Non-thermal Activity in AGN Coronae}
\label{sec:history}

\begin{figure}
 \begin{center}
  \includegraphics[width=0.7\linewidth]{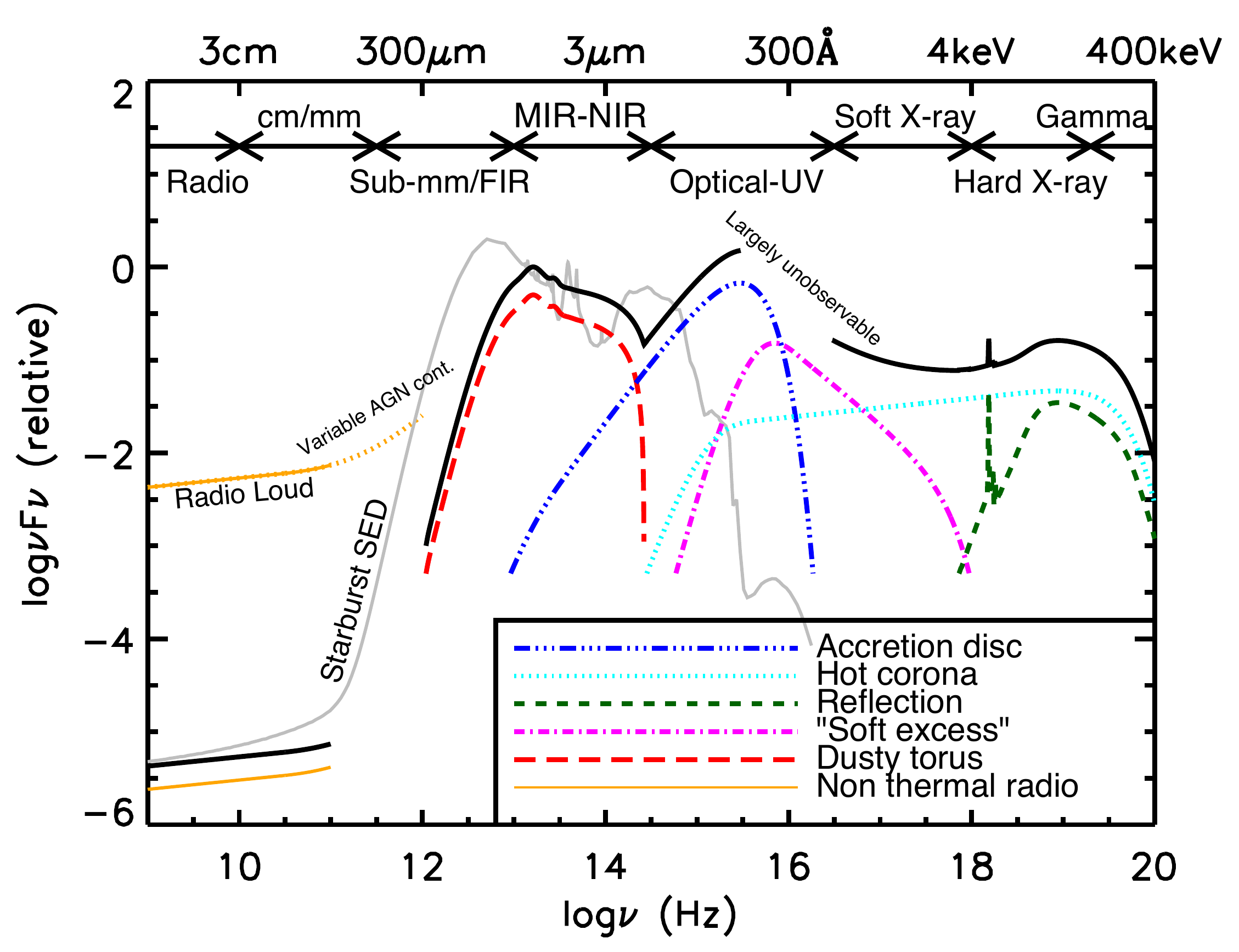}
\caption{Schematic representation of the AGN SED (black curve), separated into the main physical components, as indicated by the colored curves. For the comparison, the SED of a star-forming galaxy (light grey curve) is also shown. Taken from \citet{Harrison2014PhDT.......357H, Hickox2018ARA&A..56..625H}.}\label{fig:AGN_SED}
 \end{center}
\end{figure}

\subsection{Failure of Pair Cascade Model}
In this section, we briefly review the research history of non-thermal coronal activity in Seyferts. Scenarios involving the acceleration of high-energy particles in the nuclei of Seyferts have been discussed for a long time. 

In the 1980s, about 20 years after the dawn of X-ray astronomy, the production mechanism of X-ray emission in Seyferts was still under debate. One possibility was pair cascades induced by high energy particles \citep[e.g.,][]{Zdziarski1986, Kazanas1986, Ghisellini2004}. In the pair cascade model, particles are thought to be accelerated by shock dissipation in accretion flows \citep[e.g.,][]{Cowsik1982, Protheroe1983, Zdziarski1986, Kazanas1986, Sikora1987, Begelman1990}. These investigations tossed a coin to Seyferts as cosmic-ray factories \citep{Sikora1987, Begelman1990, Stecker1991PhRvL..66.2697S, Stecker1992}. \citet{Stecker1991PhRvL..66.2697S, Stecker1992} made first quantitative estimate of the expected neutrino fluxes from Seyferts at the level of $\sim3\times10^{-7}~{\rm GeV\ cm^{-2}\ s^{-1}\ sr^{-1}}$ at 1~PeV. However, in the 1990s, the detection of the X-ray spectral cutoffs \citep[e.g.,][]{Madejski1995, Zdziarski2000} and non-detection of Seyfert AGNs in the gamma-ray band \citep[e.g.,][]{Lin1993} ruled out the pair cascade scenario as a dominant source for X-ray emission of Seyferts, lowering the expected neutrino fluxes from Seyferts significantly. Therefore, high energy signals from AGN coronae have not been extensively investigated in the community in the late 1990s and 2000s.

\subsection{Properties of AGN Coronae Revealed by X-ray Observations}
Today, it is widely believed that the AGN X-ray emission is primarily from the Comptonized accretion disk photons from moderately thick thermal plasma, namely coronae, above an accretion disk \citep{Katz1976, 1977A&A....59..111B, Pozdniakov1977, Galeev1979, Takahara1979, Sunyaev1980}. This Comptonized emission appears together with emission reprocessed by the surrounding cold materials, a so-called Compton reflection component \citep[e.g.,][]{Lightman1988, Magdziarz1995, Ricci2011}. Figure~\ref{fig:AGN_SED} shows a current schematic representation of the AGN spectral energy distribution (SED) without obscuration \citep[see][for details]{Harrison2014PhDT.......357H, Hickox2018ARA&A..56..625H}. We note that the origin of soft X-ray excess is still under debate \citep{Done2007MNRAS.377L..59D, Boissay2016A&A...588A..70B}.

\begin{figure}
 \begin{center}
  \includegraphics[width=0.7\linewidth]{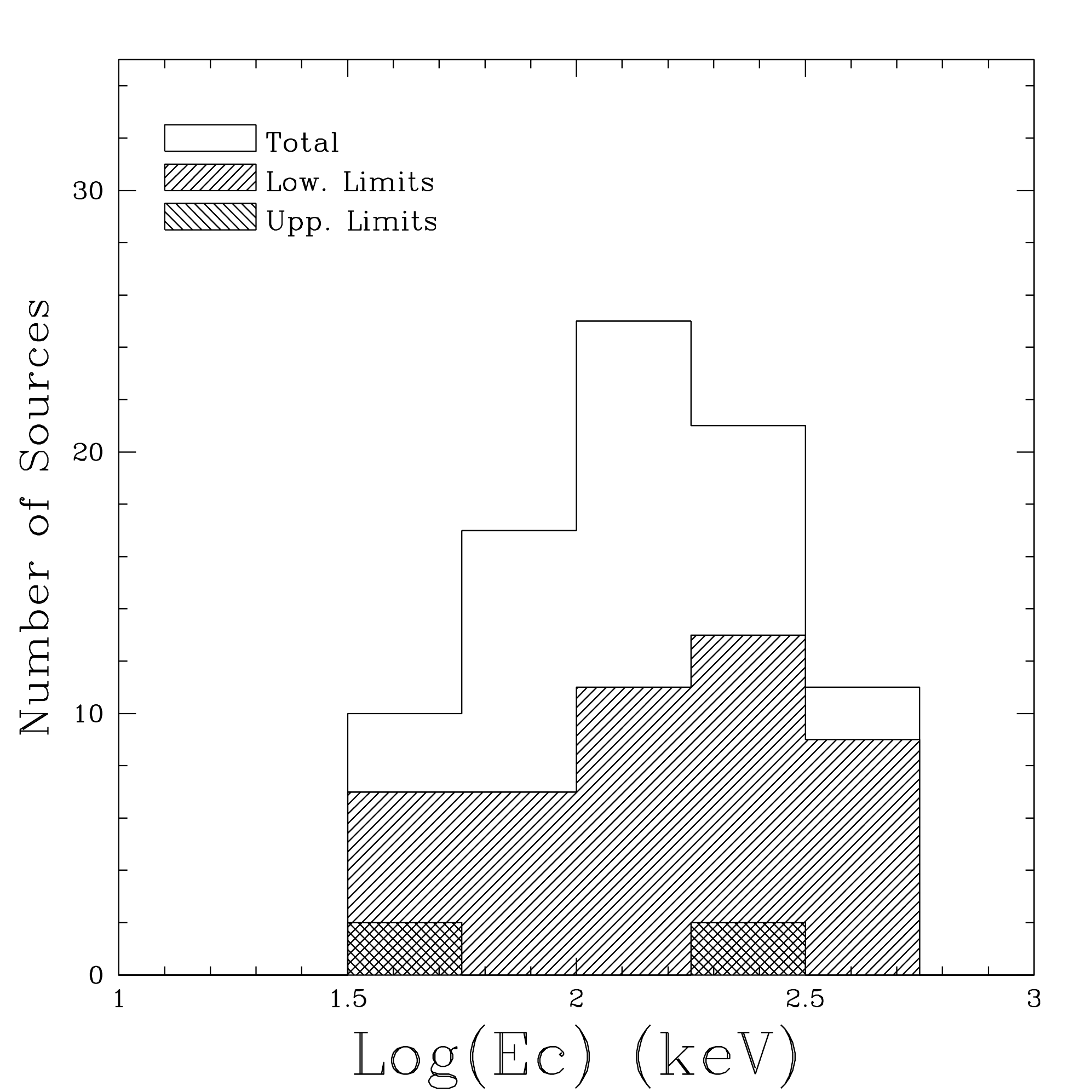}
 \end{center}
\caption{Distribution of $E_c$ of local Seyferts determined from the BeppoSAX observations. Taken from \citet{Dadina2008A&A...485..417D}.}\label{fig:E_c_Seyferts}
\end{figure}

\begin{figure}
 \begin{center}
  \includegraphics[width=0.7\linewidth]{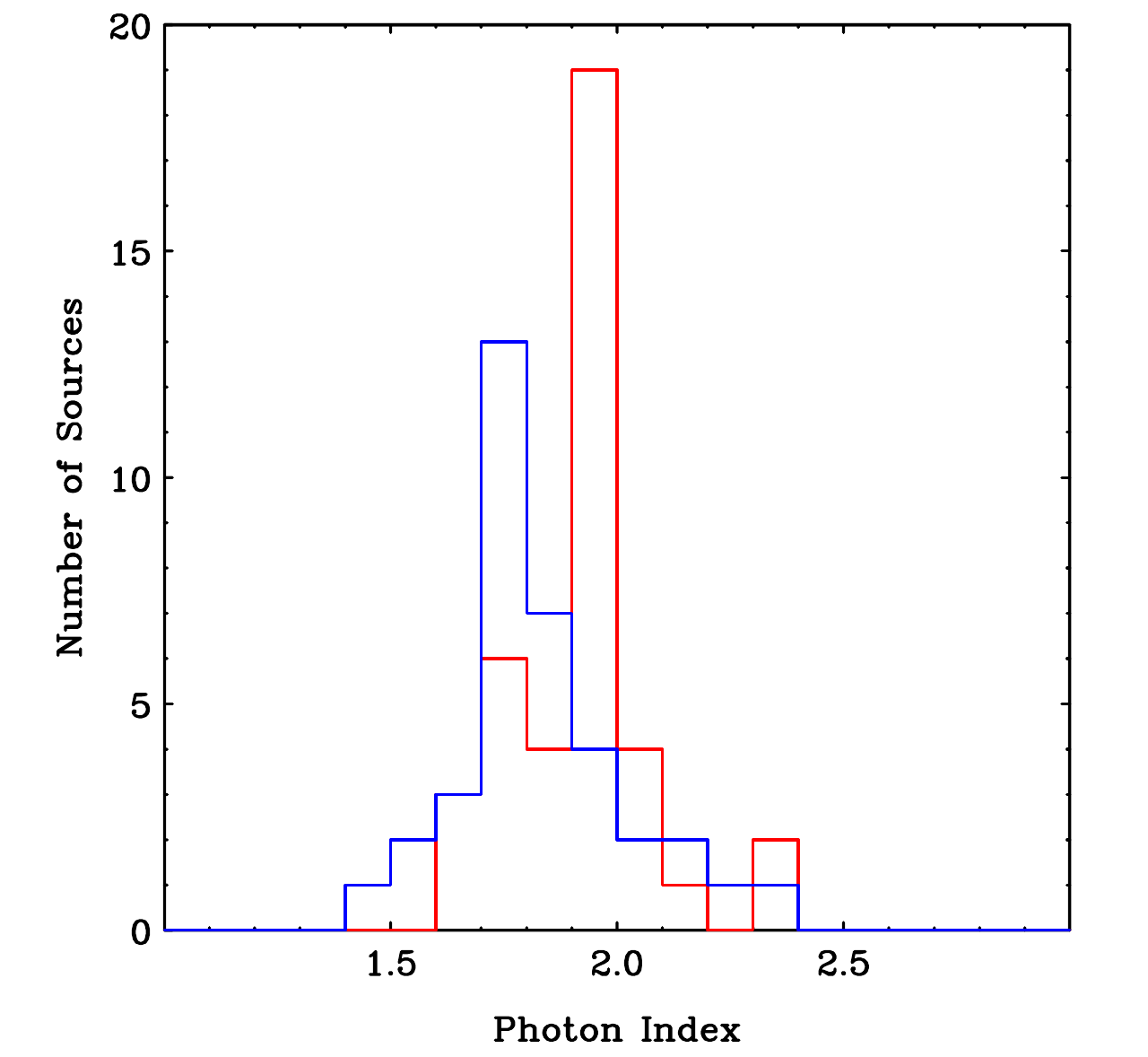} 
 \end{center}
\caption{Distribution of $\Gamma$ of local Seyferts determined from the Swift/BAT survey. Red and blue corresponds to type-1 and type-2 AGNs, respectively. Taken from \citet{Ueda2014}.}\label{fig:Gamma_Seyferts}
\end{figure}

X-ray spectral studies allow us to determine some of the coronal parameters such as the coronal electron temperature, $T_e$, and the Thomson scattering optical depth $\tau_{\rm T}\simeq n_e\sigma_{\rm T}R_c$ \citep[e.g.,][]{Brenneman2014}. Here $n_e$ is the electron number density, $\sigma_{\rm T}$ is the Thomson scattering cross section, and $R_c$ is the coronal size. \citet{Dadina2008A&A...485..417D} reported that local bright Seyferts typically have  the spectral cutoff at $E_c\approx300$~keV (Fig.~\ref{fig:E_c_Seyferts}). This cutoff corresponds to the electron temperature of $kT_e\approx100$~keV (here $k$ is the Boltzmann constant). The process of Comptonization by thermal plasma is described by the Kompaneets equation \citep{Kompaneets1957}. The photon index of the primary X-ray emission of Seyferts is typically $\Gamma\approx1.9$ \citep{Ueda2014} (Fig.~\ref{fig:Gamma_Seyferts}). This corresponds to $\tau_{\rm T}\approx1.1$ based on the solution to the Kompaneets equation \citep{Zdziarski1996} as $\Gamma = ({9}/{4}+{1}/{[\theta_e\tau_{\rm T}(1+\tau_{\rm T}/3)]})^{1/2}-{1}/{2}$, where the dimensionless electron temperature $\theta_e\equiv kT_e/m_ec^2$. It should be also mentioned that simultaneous optical--X-ray spectral fitting studies \citep{Jin2012} and microlensing observation \citep{Morgan2012} suggested the corona size $R_c\sim10R_s$, where $R_s$ is the Schwarzschild radius.

Hard X-ray observations of Seyferts allow us to investigate the AGN coronae further. It has constrained the amount of non-thermal particles in the AGN coronae from the non-detection of the non-thermal features. In order not to violate the {\it NuSTAR} observations, the latest constraint on the non-thermal electron energy fraction is obtained as $f_{\rm nth} < 0.3$ \citep{Fabian2017}.

\begin{figure}
 \begin{center}
  \includegraphics[width=0.7\linewidth]{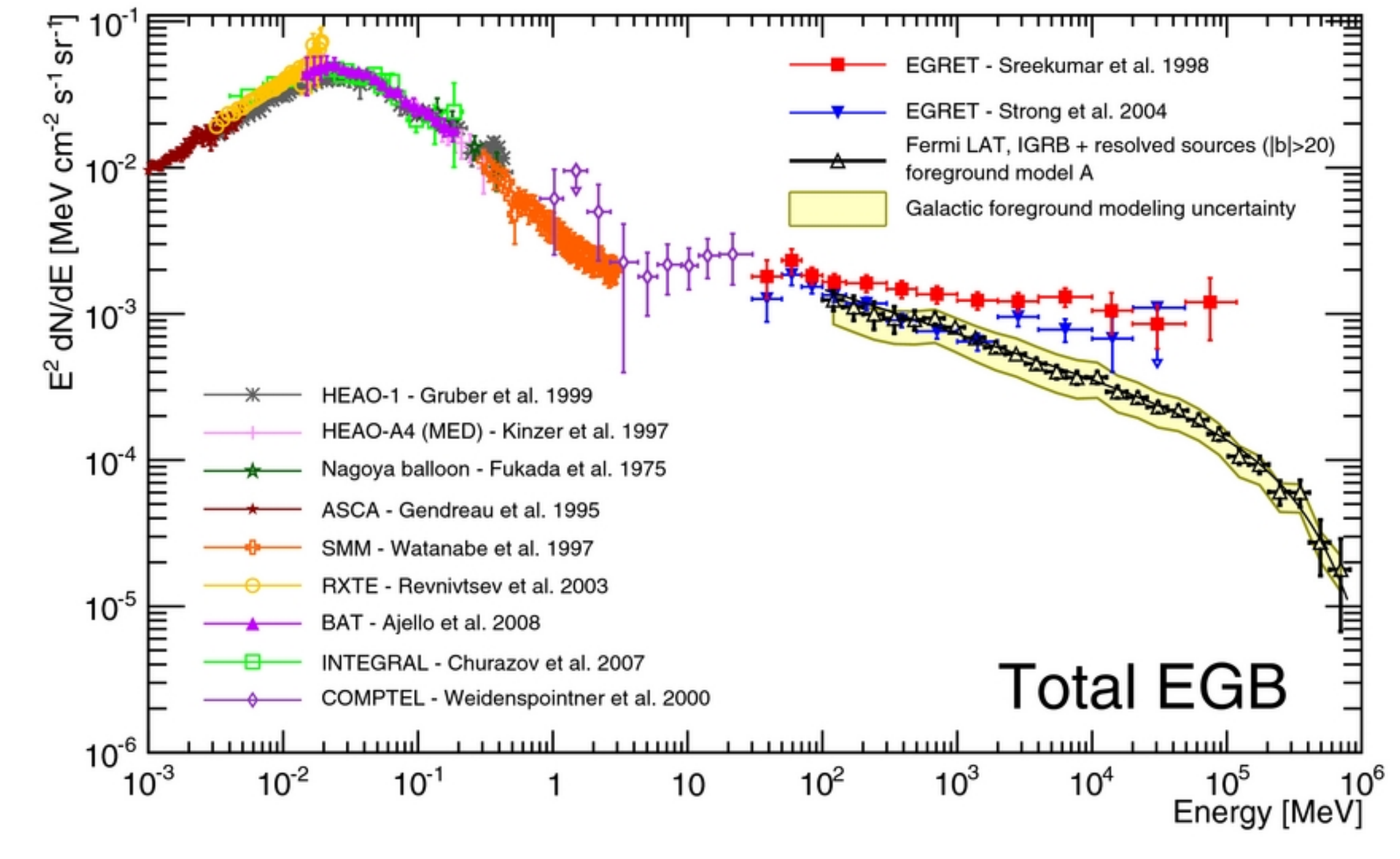}
\caption{Cosmic X-ray, MeV gamma-ray, and GeV gamma-ray background radiation spectrum. Taken from \citet{Fermi_CGB_2015ApJ...799...86A}.}\label{fig:CXGB}
 \end{center}
\end{figure}

Together with the success of the Comptonized corona model, population studies of Seyferts revealed that Seyferts dominate the cosmic X-ray background radiation up to $\sim 300$~keV \citep[e.g.,][]{Ueda2003, Hasinger2005A&A...441..417H}. Here, as shown in Fig.~\ref{fig:CXGB}, the cosmic MeV gamma-ray background radiation spectrum smoothly extends from the cosmic X-ray background spectrum \citep{Fukada1975Natur.254..398F, Watanabe1997AIPC..410.1223W, Weidenspointner2000AIPC..510..467W}. Because of a spectral cutoff in individual Seyferts, integrated Seyfert X-ray photons would not explain the cosmic MeV gamma-ray background radiation. Several theoretical models proposed that a small fraction of non-thermal electrons in AGN coronae may be enough to explain the MeV background radiation as well \citep{Stecker1999astro.ph.12106S, Inoue2008}. The required fraction was $f_{\rm nth}\sim0.03$. This amount indicated the possible existence of weak non-thermal activity in AGN coronae. However, it was an ad hoc way to have a simultaneous explanation on the X-ray and MeV backgrounds.\footnote{Blazars are another possible origin of the cosmic MeV gamma-ray background radiation \citep{Ajello2009}. However, recent studies suggest that blazars can explain only $\sim3$\% of the cosmic MeV gamma-ray background radiation \citep{Toda2020ApJ...896..172T}.}

\subsection{Properties of AGN Coronae Revealed by Millimeter Observations}
Theoretically, it is natural to expect that AGN coronae are magnetized \citep{Haardt1991, Liu2002}. Then, coronal non-thermal electrons, as expected for the MeV background radiation, inevitably generate coronal synchrotron emission \citep{DiMatteo1997, Inoue2014, Raginski2016}, which should appear as an excess in the millimeter (mm) band. Inconclusive signs of such a new component in the radio spectra of several Seyfert galaxies have been reported in literature \citep{Antonucci1988, Barvainis1996, Doi2016, Behar2018}. However, a paucity of multi-band data and the contamination of extended dust emission hamper the investigation. 

\begin{figure*}[t]
 \begin{center}
  \includegraphics[width=0.7\linewidth]{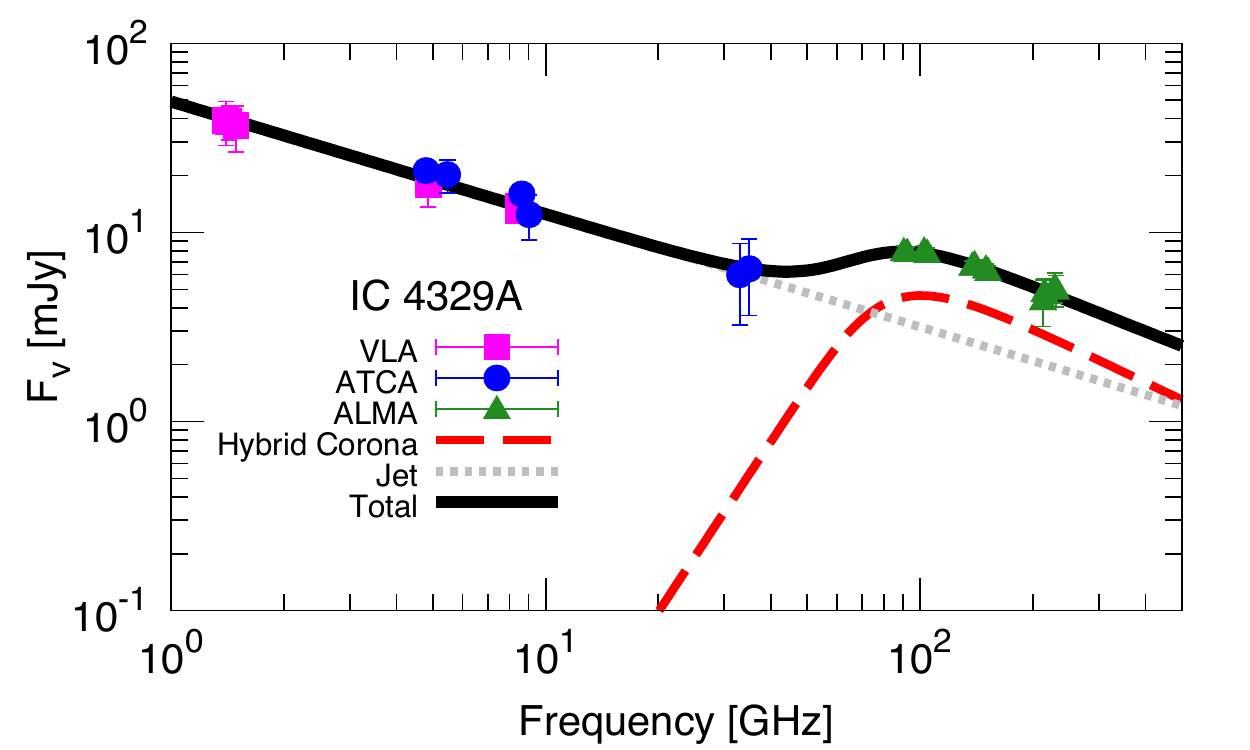} 
 \end{center}
\caption{The cm-mm spectrum of IC~4329A after subtracting extended emission due to galactic star formation activity. The square, circle, and triangle points show the VLA, ATCA, and ALMA data, respectively. The error bars correspond to 1-$\sigma$ uncertainties. The dashed and dotted lines show the fitted hybrid corona and jet component, respectively. The solid line shows the sum of these two components. Taken from \citet{Inoue2018}.}\label{fig:SED_radio}
\end{figure*}

Recently, \citet{Inoue2018} reported the detection of power-law coronal radio synchrotron emission from two nearby Seyferts, IC~4329A and NGC~985, utilizing ALMA (See Figs.~\ref{fig:SED_radio}), which enabled multi-band observations with high enough angular resolution to exclude the galactic contamination. These observations provided the first determination of the fundamental physical parameters of the AGN coronae: magnetic field strength and its size. The inferred coronal magnetic field strength $B_c$ was $\sim10$~G with a size $R_c$ of $40R_s$ for both Seyferts with a central black hole mass of $\sim10^8M_\odot$. 

\begin{figure}[h]
 \begin{center}
  \includegraphics[width=0.7\linewidth]{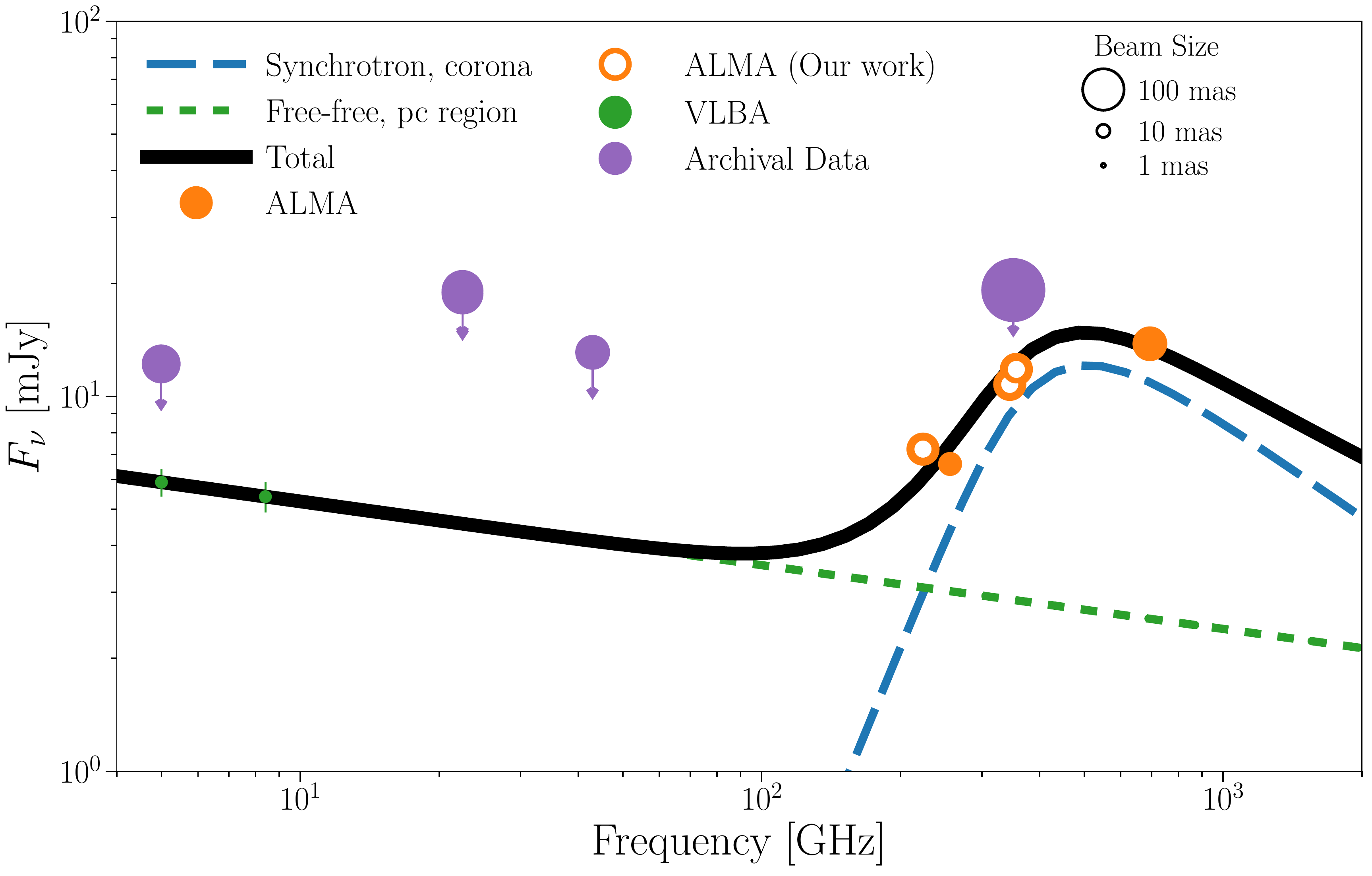}
\caption{The cm-mm spectrum of NGC~1068. The data points from VLBA \citep{Gallimore2004} and ALMA \citep{GarciaBurillo2016, Impellizzeri2019, GarciaBurillo2019} are shown in green and orange, respectively. The open points represent the newly analyzed ALMA data. The size of circles corresponds to the beam sizes as indicated in the figure. We also show the archival mm-cm data having large beam sizes as upper limits in purple \citep{Gallimore1996, Cotton2008, Pasetto2019}. The error bars correspond to 1-$\sigma$ uncertainties. The blue-dashed and green-dotted lines show the coronal synchrotron and {pc-scale} free-free component, respectively. The solid black line shows the sum of these two components. Taken from \citet{Inoue2020ApJ...891L..33I}.}\label{fig:ALMA}
 \end{center}
\end{figure}

Coronal synchrotron emission is also reported for NGC~1068 in \citet{Inoue2020ApJ...891L..33I}. Fig. \ref{fig:ALMA} shows the cm-mm spectrum of NGC~1068, which shows a mm-excess similar to IC~4329A and NGC~985. The coronal synchrotron emission can reproduce the mm excess of NGC~1068 with parameters of $B_c=100$~G, $R_c=10~R_s$, and the spectral index of non-thermal electrons $p=2.7$.

Power-law coronal synchrotron emission also suggests that AGN coronae contain both thermal and non-thermal electrons, i.e., hybrid coronae. Therefore, acceleration of high-energy particles should happen in AGN coronae. However, the exact $f_{\rm nth}$ can not be determined by the current measurements because of degeneracy among $R_c$, $B_c$, and $f_{\rm nth}$. Here, significantly low $f_{\rm nth}\ll10^{-3}$ requires a relatively large $R_c$ based on the radio spectral fitting, which contradicts with optical--X-ray spectral fitting studies \citep{Jin2012} and microlensing observations \citep{Morgan2012}. Thus, we may have weak but non-negligible non-thermal coronal activity at the level of $10^{-3}<f_{\rm nth}<0.3$.

\citet{Inoue2018} also suggested that the coronae are likely to be advection heated hot accretion flows \citep{Kato2008, Yuan2014} rather than magnetically heated coronae \citep{Haardt1991, Liu2002} because the measured magnetic field strength (i) is too weak to keep the coronae hot and (ii) is consistent with the value based on the self-similar solutions of hot accretion flows. Thus, we may assume that coronal magnetic field strength scales as $B\propto M_{\rm BH}^{-1/2}$ \citep{Yuan2014}, where we ignore its dependence on the accretion rate and other parameters for simplicity. 

\section{Generation of High Energy Particles in AGN Coronae}
\label{sec:model}
Previous non-thermal coronal models have overestimated the non-thermal activity strength since various coronal parameters are not well determined. As described above, current X-ray and radio observations now allow us to determine the corona size $R_c$, the electron density $n_e$, the magnetic field $B_c$, non-thermal electrons' spectral index $p$, and even the non-thermal electron energy fraction $f_{\rm nth}$. With that knowledge, we can investigate the high-energy particle production processes in AGN coronae again. This section overviews the energy loss and acceleration processes of high-energy particles in the AGN coronae.

The geometry of black hole coronae is still under debate. A recent polarization study revealed the extended coronal structure in a nearby black hole binary \citep{Chauvin2018NatAs...2..652C}. Such extended coronal geometry is also naturally expected in AGNs. The detailed structure of extended coronae can be a spherical, slab, or patchy. For simplicity, in this Review, coronae are assumed to be spherical with a radius of $R_c\equiv r_cR_s$, where $r_c$ is the dimensionless corona size. 

Coronae are also set to be in a steady state. The proton number density $n_p$ is set to be equal to $n_e$, which gives the maximum number of protons in coronae. $n_e$ is defined through $\tau_{\rm T}$  as 
\begin{equation}
	n_e = \frac{\tau_{\rm T}}{\sigma_{\rm T} R_c}\simeq 1.4\times10^9\left(\frac{\tau_{\rm T}}{1.1}\right)\left(\frac{r_c}{40}\right)^{-1}\left(\frac{M_{\rm BH}}{10^8M_\odot}\right)^{-1}\ {\rm cm}^{-3}.
\end{equation}

For simplicity, the gas is assumed to be accreted on to the SMBH with free-fall velocity $v_{\rm ff} = \sqrt{2GM_{\rm BH}/R_c}$. The free-fall timescale from the coronal region is estimated to be
\begin{equation}
\label{eq:t_fall}
t_{\rm fall}= R_c / v_{\rm ff}\simeq2.5\times10^5\left(\frac{r_c}{40}\right)^{3/2}\left(\frac{M_{\rm BH}}{10^8M_\odot}\right)\ [{\rm s}].
\end{equation}

\subsection{Energy Loss Processes}
High energy particles loose their energies through radiative cooling processes. In AGN coronae, high-energy electrons mainly lose their energies via synchrotron and inverse Compton (IC) radiation. The synchrotron cooling rate for an electron with a Lorentz factor of $\gamma_e$ is
\begin{equation}
 t_{{\rm syn}, e}(\gamma_e) = \frac{3}{4} \frac{m_e c}{\sigma_{\rm T}  U_{\rm B}} \gamma_e^{-1}\simeq 7.7\times10^4\left(\frac{B_c}{10~{\rm G}}\right)^{-2}\left(\frac{\gamma_e}{100}\right)^{-1}\ [{\rm s}],
\end{equation}
where $m_e$ is the electron rest mass and $U_{\rm B} =B_c^2/8\pi$ is the magnetic field energy density of magnetic field strength $B$.

The inverse Compton cooling rate including the Klein-Nishina cross section \citep{Jones1968,Moderski2005,Khangulyan2014} is
\begin{equation}
\label{eq:time_ic}
t_{\rm IC}(\gamma_e) = \frac{3 m_e c}{4\sigma_{\rm T}}\left[\int_0^{\infty}d\epsilon f_{\rm KN}(\tilde{b})\frac{U_{\rm ph}(\epsilon)}{\epsilon} \right]^{-1}\gamma_e^{-1},
\end{equation}
where $\tilde{b}\equiv 4\gamma_e\epsilon/m_ec^2$ and $f_{\rm KN} \simeq 1/(1.0+\tilde{b})$ \citep{Moderski2005}. The photon energy density, $U_{\rm ph}$, is given as $U_{\rm ph}(\epsilon)=L_{\rm ph}(\epsilon)/4\pi R_c^2c$, where $L_{\rm ph}$ is the AGN core (disk + corona) luminosity and  $\epsilon$ is the photon energy.

Relativistic protons are predominately cooled through inelastic $pp$ interactions and $p\gamma$ reactions. Proton synchrotron and IC cooling channels are inefficient as compared to the hardronic mechanisms. Hereinafter, we do not consider proton IC/synchrotron coolings. The $pp$ cooling time can be expressed as
\begin{equation}
\label{eq:t_pp}
 t_{pp} = \frac{1}{n_p\sigma_{pp} c \kappa_{pp}}\simeq 1.6\times10^6\left(\frac{\tau_{\rm T}}{1.1}\right)^{-1}\left(\frac{r_c}{40}\right)\left(\frac{M_{\rm BH}}{10^8M_\odot}\right)\ [{\rm s}].
\end{equation}
where $\kappa_{pp}\sim 0.5$ is the proton inelasticity of the process and we adopt $\sigma_{pp}=3\times10^{-26}\ {\rm cm}^2$ and \(n_p\approx n_e\).  Below we follow the formalism developed by \citet{Kelner2006}.

The $p\gamma$ cooling time via photomeson interactions is
\begin{equation}
\label{eq:t_pg}
 t_{p\gamma}^{-1} = \frac{c}{2\gamma_p^2}
  \int_{\bar{\varepsilon}_{\rm thr}}^{\infty}d\bar{\varepsilon}\sigma_{p\gamma}(\bar{\varepsilon})K_{p\gamma}(\bar{\varepsilon})\bar{\varepsilon} \int_{\bar{\varepsilon}/(2\gamma_p)}^{\infty}d\epsilon\, \frac{U_{\rm ph}(\epsilon)}{\epsilon^4},
\end{equation}
where $\gamma_p$ is the proton Lorentz factor, and $\bar{\varepsilon}$ and $\epsilon$ are the photon energy in the proton rest frame and the black hole frame, respectively, 
$U_{\rm ph}$ is the energy density of the photon target, and $\bar{\varepsilon}_{\rm thr} = 145$~MeV \citep{Kelner2008}.

The $p\gamma$ interaction also generates secondary leptons and enable  pair production via the so-called Bethe-Heitler process. Cooling timescale for Bethe-Heitler process is approximated as \citep{Gao2012}
\begin{equation}
t_{\rm BH}^{-1} \approx\frac{7(m_{e}c^{2})^{3}\alpha_{f}\sigma_{\rm T} c}{9\sqrt{2}{\pi}m_{p}c^2\gamma_{p}^{2}}\int_{m_ec^2/\gamma_p}^{\infty}d\epsilon\frac{U_{\rm ph}(\epsilon)}{\epsilon^4} \left\{\left(\frac{2\gamma_{p}\epsilon}{m_ec^2}\right)^{3/2}\left[\log\left(\frac{2\gamma_{p}\epsilon}{m_ec^2}\right) -2/3\right]+2/3\right\},
\end{equation}
where $m_p$ is the proton rest mass and $\alpha_f$ is the fine-structure constant.

\subsection{Acceleration}

Various acceleration mechanisms can take place in the coronae such as diffusive shock acceleration (DSA) mechanism \cite[e.g.,][]{Drury1983,Blandford1987}, turbulent acceleration \citep[e.g.,][]{Zhdankin2018}, magnetosphere acceleration \citep[e.g.,][]{Beskin1992,Levinson2000}, and magnetic reconnection \citep[e.g.,][]{Hoshino2012}. In this section, we consider these processes given our current observational knowledge (\S.~\ref{sec:history}).

\subsubsection{Diffusive Shock Acceleration}
In the frame work of DSA \cite[e.g.,][]{Drury1983,Blandford1987}, the acceleration time scale can be approximated as 
\begin{equation}
t_{\rm DSA}\simeq\frac{\eta_{\rm acc}D(E_{\rm CR})}{v_{\rm sh}^2},
\end{equation}
where $D$ is the diffusion coefficient, $E_{\rm CR}$ is the particle energy, and $v_{\rm sh}$ is the shock speed. $\eta_{\rm acc}$ is a numerical factor that depends on the shock compression ratio and the spatial dependence of $D$ \citep{Drury1983}. We set $\eta_{\rm acc}=10$. Assuming a Bohm-like diffusion, 
\begin{equation}
D(E_{\rm CR} )\simeq\frac{\eta_gcE_{\rm CR} }{3eB_c},
\end{equation}
where $e$ is the electric charge and $\eta_g$ is the gyrofactor which is the mean free path of a particle in units of the gyroradius. $\eta_g$ characterizes the efficiency of the acceleration. $\eta_g=1$ corresponds to the Bohm limit case. The DSA time can be written as
\begin{equation}
\label{eq:t_acc}
t_{\rm DSA}\simeq\frac{10}{3}\frac{\eta_g c R_g}{v_{\rm sh}^2} \simeq 7.6\times10^{-3} \left(\frac{\eta_g}{100}\right)\left(\frac{m_{p/e}}{m_e}\right)\left(\frac{r_c}{40}\right)\left(\frac{B_c}{10\ {\rm G}}\right)^{-1}\left(\frac{\gamma_{p/e}}{100}\right)\ [{\rm s}].
\end{equation}
where $R_g=E_{\rm CR}/eB_c$ is the gyro radius. $v_{\rm sh}$ is set as $v_{\rm ff}$. Thus, $r_c$ appears in Eq.~\ref{eq:t_acc}. $\eta_g$ varies in different astrophysical environments. $\eta_g\sim1$ is possibly seen in a Galactic supernova remnant \citep{Uchiyama2007} and a microquasar jet \citep{Sudoh2020ApJ...889..146S}, while $\eta_g\sim10^4$ is seen in the case of blazars in the framework of one-zone leptonic models \citep[e.g.,][]{Inoue1996,Finke2008,Inoue2016}.

\subsubsection{Stochastic Acceleration}
Stochastic (turbulent) acceleration has been considered for low-accretion rate objects such as low-luminosity AGN \citep[e.g.,][]{Kimura2015,Zhdankin2017,Zhdankin2018,Wong2019}. In this scenario, particles are accelerated stochastically by turbulence and magnetic reconnection in accretion disk or coronae. We briefly follow the stochastic acceleration in the AGN coronae case. According to the quasi-linear theory, the diffusion coefficient in the momentum space is \cite[e.g.,][]{Dermer1996}
\begin{equation}
D_p\simeq (m_pc)^2 (ck_{\rm min})\left(\frac{v_A}{c}\right)^2\zeta(r_L k_{\rm min})^{q-2}\gamma^q,
\end{equation}
where $k_{\rm min}\sim R_c^{-1}$ is the minimum wave number of turbulence spectrum (corresponding to the size of the corona), $v_A =B_c/\sqrt{4\pi m_p n_p}$ is the Alfv\'en speed, $r_g=m_pc^2/eB_c$ is the gyro radius, $\zeta=\delta B_c^2/B_c^2$ is the ratio of strength of turbulence fields against the background, and $q$ describes the spectrum of the turbulence. Then, the acceleration timescale is estimated to be
\begin{equation}
t_{\rm st}\simeq\frac{p^2}{D_p}\simeq\frac{1}{\zeta}\left(\frac{v_A }{c}\right)^{-2}\frac{R_c}{c}\left(\frac{r_L}{R_c}\right)^{2-q}\gamma^{2-q}
\end{equation}

Assuming the Kolomogorov spectrum for the turbulence ($q=5/3$) and $\zeta=1$, the timescale becomes
\begin{equation}
t_{\rm st}\simeq 3.1\times10^7\left(\frac{\tau_{\rm T}}{1.1}\right)\left(\frac{r_c}{40}\right)^{-1/3}\left(\frac{M_{\rm BH}}{10^8M_\odot}\right)^{-1/3}\left(\frac{B_c}{10\ {\rm G}}\right)^{-7/3}\left(\frac{\gamma_p}{100}\right)^{1/3}\ [{\rm s}].
\end{equation}
Thus, stochastic acceleration appears to be inefficient compared to the typical cooling rates because of the measured weak magnetic fields, which results in slow Alfv\'en speed.

\subsubsection{Magnetosphere Acceleration}
Magnetosphere acceleration can also accelerate particles in the vicinity of SMBH \citep[e.g.,][]{Beskin1992,Levinson2000,Neronov2007,Levinson2011,Rieger2011}. At low accretion rates, the injection of charges into the BH magnetosphere is not sufficient for a full screening of the electric field induced by the rotation of the compact object. The regions with the unscreened electric field, so-called gaps, can accelerate charged particles effectively. 

In order to have gaps, the maximum allowed accretion rate is \citep{Levinson2011,Aleksic2014,Aharonian2017}
\begin{equation}
\dot{m}<3\times10^{-4}\left(\frac{M_{\rm BH}}{10^8M_\odot}\right)^{-1/7},
\end{equation}
where $\dot{m}$ is the accretion rate in the Eddington units. Since we are considering the standard accretion disk regime $\dot{m}\gtrsim0.01$, particle acceleration by gaps does not operate in our case.

\subsubsection{Reconnection Acceleration}
Magnetic reconnection would accelerate particles \citep[see e.g.,][for reviews]{Hoshino2012}. Reconnection would naturally happen in magnetized coronae, and radiative magnetic reconnection is also suggested as a possible origin of the X-ray emission seen in accreting black hole systems \citep{Beloborodov2017}. However, even in the case of solar flares, particle acceleration mechanisms in magnetic reconnection are still uncertain \citep[e.g.,][]{Liu2008, Nishizuka2013}. Therefore, a quantitative discussion is not straightforward. 

Here, given the magnetic field measurements, we can estimate the available energy injection power by global magnetic activity as 
\begin{equation}
	P_B = \frac{B_c^2R_c^2v_A}{2}\simeq 4.0\times10^{39}\left(\frac{\tau_{\rm T}}{1.1}\right)^{-1/2}\left(\frac{r_c}{40}\right)^{5/2}\left(\frac{M_{\rm BH}}{10^8M_\odot}\right)^{5/2}\left(\frac{B_c}{10~{\rm G}}\right)^3\ [{\rm erg\ s^{-1}}].
\end{equation}
This power seems not sufficient for providing the non-thermal particle energies under one-zone estimates. However, highly non-homogeneous configurations of magnetic field, such as by a local magneto-rotational instability \citep{Balbus1991}, may provide enough energy to produce non-thermal particles.

\begin{figure*}[tb!]
 \begin{center}
  \includegraphics[width=0.47\linewidth]{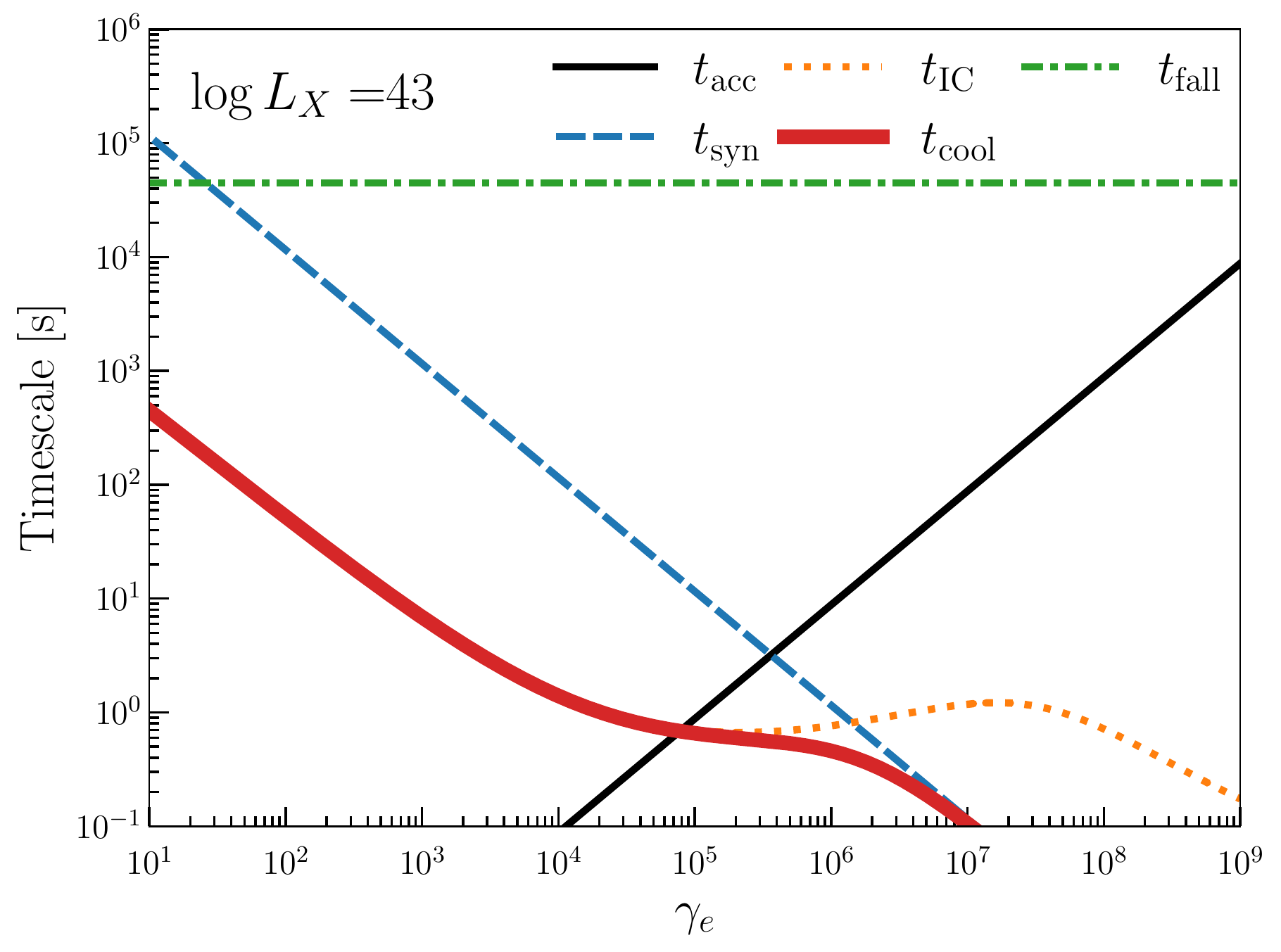}
  \includegraphics[width=0.47\linewidth]{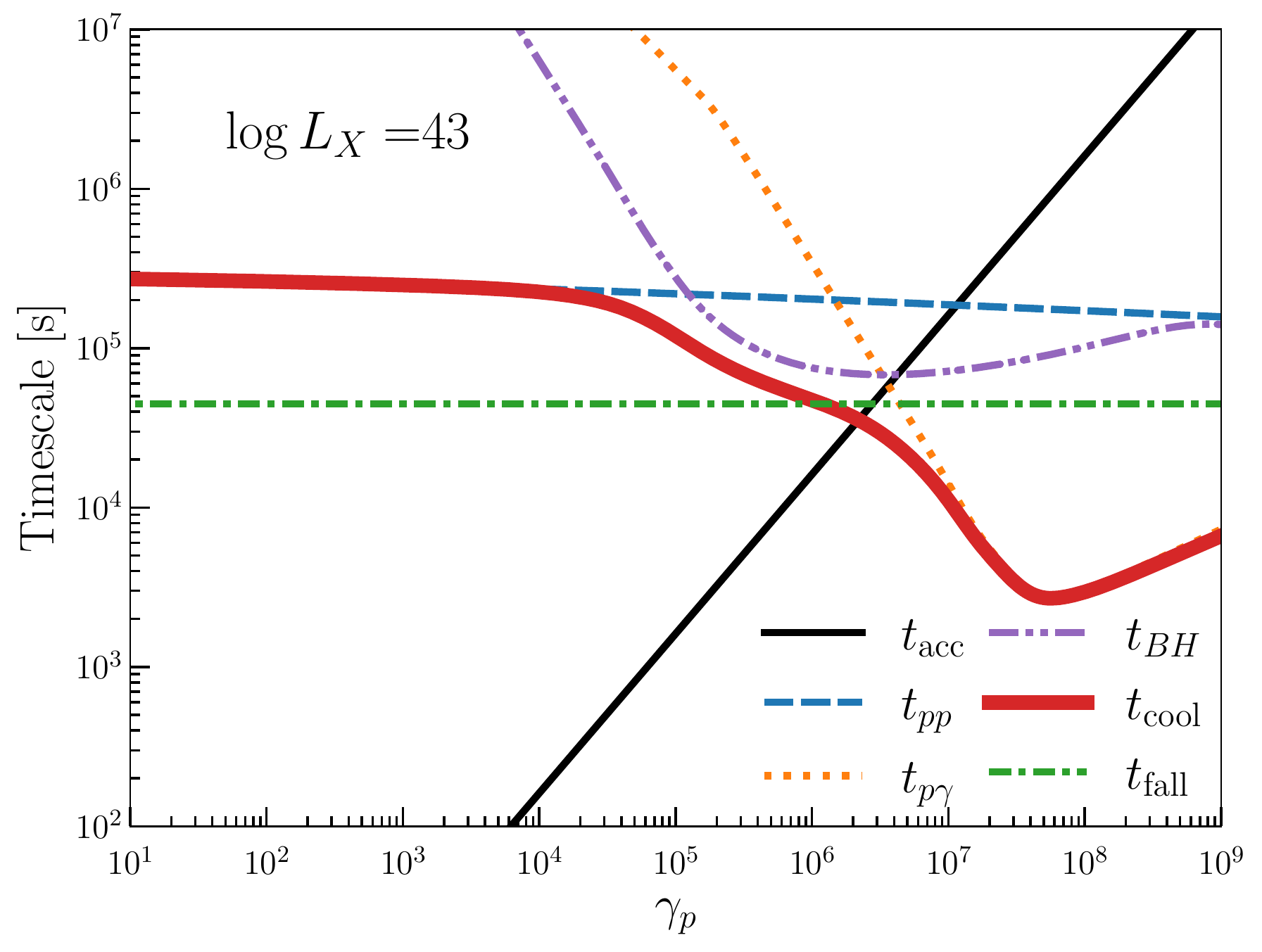}
 \end{center}
\caption{{\it Left}: Electron energy losses in AGN coronae together with acceleration and dynamical timescales for $L_X=10^{43}\ {\rm erg\ s^{-1}}$. Thin solid line shows the acceleration timescale assuming DSA. Dashed, dotted, and thick solid curve corresponds to synchrotron cooling, IC cooling, and total cooling timescale, respectively. Dot-dashed curve shows the free-fall timescale. {\it Right}: Same as in the {\it Left} panel, but for protons. Dashed, dotted, double-dot-dashed, and thick solid curve corresponds to $pp$ cooling,  $p\gamma$ cooling, Bethe-Heitler cooling, and total cooling timescale, respectively. For both panels, different luminosity cases are shown in \citet{Inoue2019ApJ...880...40I}.}\label{fig:time_CR}
\end{figure*}
\subsection{Comparison of Timescales}

Given the observed properties of AGN core regions, one can estimate the radiative cooling and acceleration timescales of high-energy particles in the coronae. Here, we take the DSA as a fiducial acceleration process. 

Left panel of Fig.~\ref{fig:time_CR} shows the timescales of high energy electrons for $\log L_X=43$ \citep[see][for details]{Inoue2019ApJ...880...40I} With $\eta_g=30$, electron acceleration up to $\gamma_e\sim10^5$ ($\sim50$~GeV) is feasible in AGN coronae. Because of the intense radiation field, Compton cooling dominates the cooling. However, at higher energy regions, the main cooling channel is replaced by synchrotron cooling due to the Klein-Nishina effect. We note that the dominance of photon fields over the magnetic field does not necessarily prevent particle acceleration as such conditions are met in some efficient non-thermal sources, e.g., in gamma-ray binary systems \citep{2006JPhCS..39..408A,2008MNRAS.383..467K}. Moreover, the high density of target photons can enable the converter acceleration mechanism if a relativistic velocity jump present in the system \citep{2003PhRvD..68d3003D}. 

Right panel of Fig.~\ref{fig:time_CR} shows the timescales for protons \citep[see][for details]{Inoue2019ApJ...880...40I}. Protons can be accelerated up to $\gamma_p\sim10^6$ ($\sim1$~PeV) in AGN coronae. Maximum attainable energy is controlled by different processes for different luminosity AGNs. For low-luminosity Seyferts ($L_X<10^{44}\ {\rm erg\ s^{-1}}$), acceleration is limited by the dynamical timescale rather than radiative cooling, while it becomes limited by the Bethe-Heitler cooling for higher luminosity objects.

\subsection{Particle Spectrum}
\label{sec:particle_spectrum}

\begin{figure}
 \begin{center}
  \includegraphics[width=0.7\linewidth]{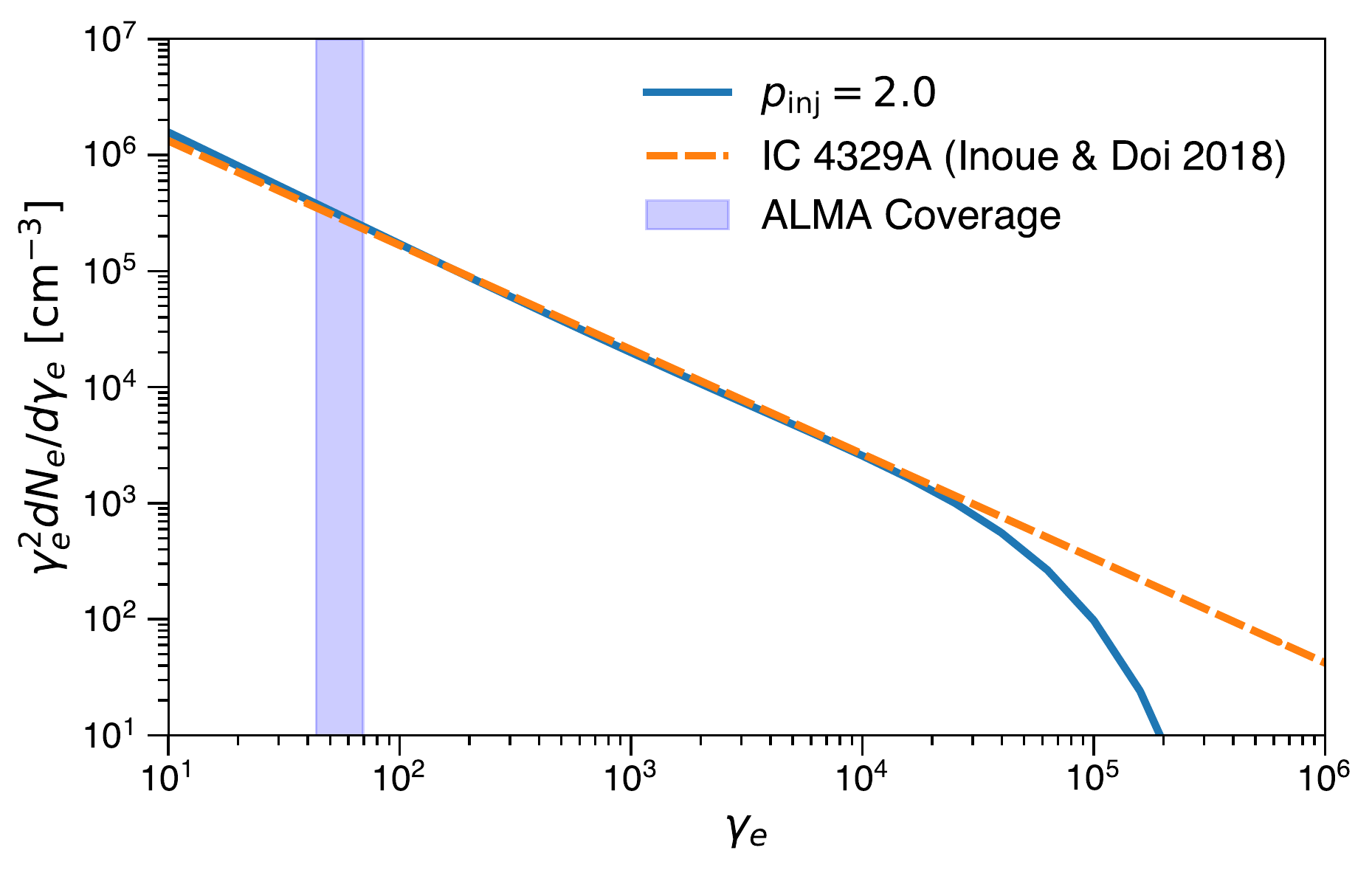}
\caption{The steady-state electron spectral distribution in AGN coronae. Solid curve corresponds to the model with $p_{\rm inj}=2.0$. Dashed curve corresponds to the observationally determined electron distribution for IC~4329A \citep{Inoue2018}. The shaded region shows the Lorentz factors responsible for the observed radio spectrum.}\label{fig:electron_spectrum}
 \end{center}
\end{figure}

The steady state particle distributions $n=dN/d\gamma$ can be derived from the solution of the transport equation \citep{Ginzburg1964}
\begin{equation}
    \frac{\partial}{\partial \gamma}\left(\dot{\gamma}_{\rm cool}n\right)+\frac{n}{t_{\rm fall}} = Q(\gamma),
\end{equation}
where $\dot{\gamma}_{\rm cool}$ is the total cooling rate, $Q(\gamma)$ is the injection function, which describes phenomenologically some acceleration process. $Q(\gamma)$ is set as $Q_0\gamma^{-p_{\rm inj}}\exp(-\gamma/\gamma_{\rm max})$. Here, $\gamma_{\rm max}$ is the maximum Lorentz factor determined by balancing the acceleration and cooling time scales.  The corresponding solution is
\begin{equation}
\label{eq:electron_spectrum}
    n=\frac{1}{|\dot{\gamma}_{\rm cool}|}\int\limits_\gamma^{\infty}Q(\gamma')e^{-T(\gamma,\gamma')} d\gamma',
\end{equation}
where
\begin{equation}
    T(\gamma_1, \gamma_2) = \frac{1}{t_{\rm fall}}\int\limits_{\gamma_1}^{\gamma_2}\frac{d\gamma}{|\dot{\gamma}_{\rm cool}|}
\end{equation}
By solving Equation. \ref{eq:electron_spectrum}, we can obtain a steady-state spectrum of the non-thermal particles. Fig.~\ref{fig:electron_spectrum} shows the steady-state non-thermal electron spectrum obtained for the injection spectral index of $p_{\rm inj}=2.0$ together with the observationally determined electron spectral distribution for IC~4329A \citep{Inoue2018}. 
The synthetic electron distribution obtained for $p_{\rm inj}=2.0$ nicely reproduces the observationally determined electron spectrum in the energy range constrained by the observations. This injection index is naturally expected in a simple DSA scenario for a strong shock.

\subsection{Energy Injection}
The total shock power $P_{\rm sh}$ can be estimated as \begin{equation}
	P_{\rm sh}=2\pi R_c^2 n_pm_p v^3_{\rm fall}\simeq2.2\times10^{45}\left(\frac{\tau_{\rm T}}{1.1}\right)\left(\frac{r_c}{40}\right)^{-1/2}\left(\frac{M_{\rm BH}}{10^8M_\odot}\right)\ {\rm erg\ s^{-1}}.
\end{equation}
$f_{\rm nth}=0.03$ corresponds to $\sim5$\% of the shock power is injected into the acceleration of electrons. Moreover, to explain the observed IceCube neutrino fluxes, the same energy injection rate is achieved for protons \citep{Inoue2019ApJ...880...40I}. This high value implies that if DSA is responsible for particle acceleration in AGN coronae, then processes regulating the injection of electrons into DSA are very efficient. For example, in the case of DSA in supernovae remnants, non-thermal electrons obtain only $\sim1$\% of energy transferred to non-thermal protons \citep{Ackermann2013}. Detailed consideration of the reasons for this unusually high efficiency of electron acceleration is needed to be studied in the future. However, recent particle-in-cell simulations of proton-electron plasma considering radiatively inefficient accretion flows showed that the energy ratio depends on the proton temperature. Higher proton temperature will result in higher electron energy fraction \citep{Zhdankin2018}. A further detailed investigation in the corona cases is required.

\begin{figure}
 \begin{center}
  \includegraphics[width=0.7\linewidth]{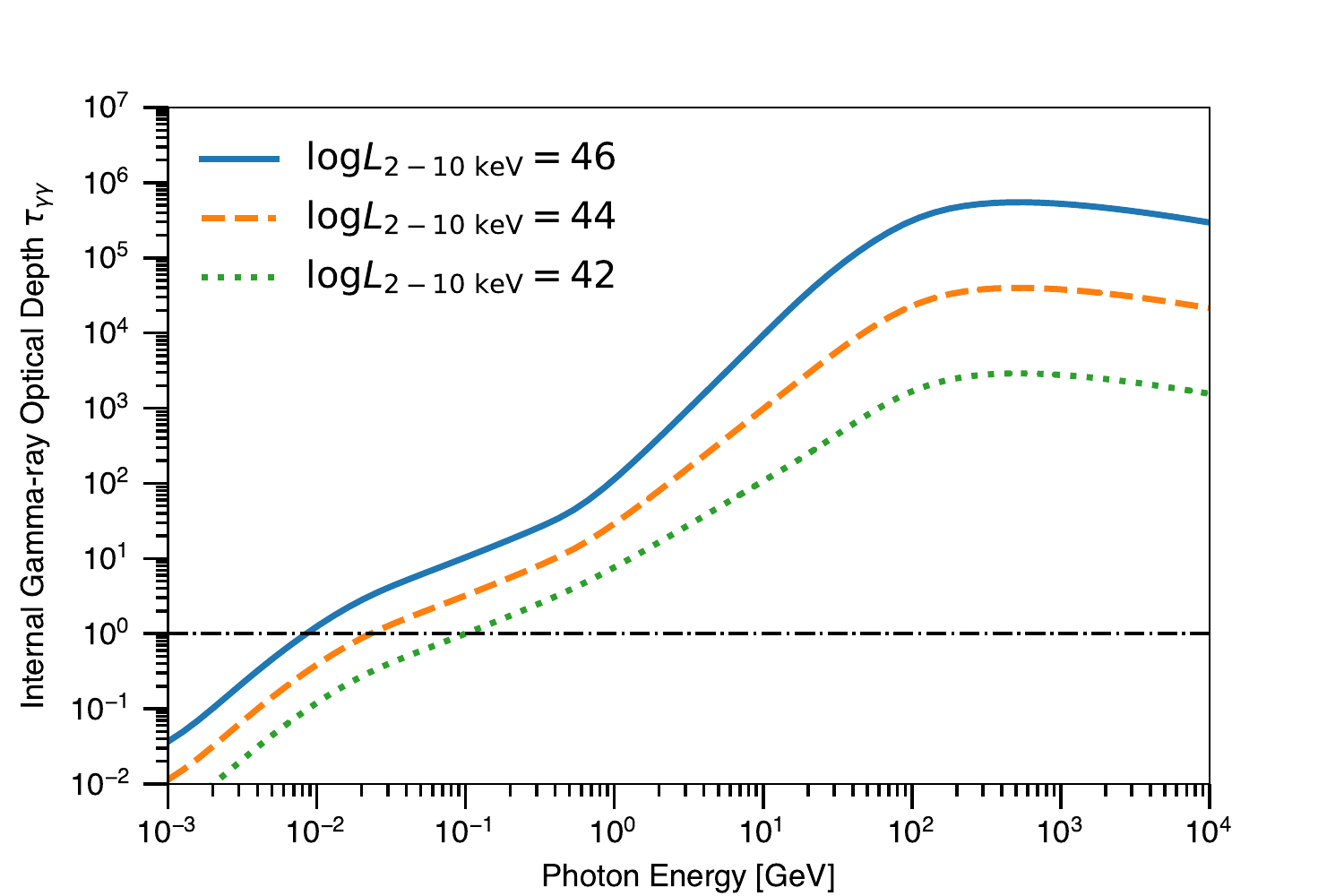}
\caption{Internal gamma-ray optical depth in the core region of AGNs. From top to bottom, each curve corresponds to 2-10~keV luminosity of $10^{46}$, $10^{44}$, $10^{42}~{\rm erg\ s^{-1}}$, respectively. The horizontal dot-dashed line represents $\tau_{\gamma\gamma}=1$.}\label{fig:tau_gg}
 \end{center}
\end{figure}

\section{Gamma Rays and Neutrinos from AGN Coronae}
\label{sec:g_nu_AGN}
\subsection{Internal Gamma-ray Attenuation in Coronae}
\label{subsec:gg_int}
Accelerated electrons and protons in AGN coronae generate gamma-ray and neutrino emission through IC scattering, $pp$ interaction, and $p\gamma$ interaction. Those high energy gamma-ray photons are attenuated by photon-photon pair production interactions ($\gamma\gamma \rightarrow e^+e^-$) with low-energy photons \citep{1934PhRv...46.1087B,Heitler1954,Aharonian_book}. We can compute the optical depth for high-energy gamma rays to $\gamma\gamma$ pair production interactions from the SED of AGN core regions. Figure~\ref{fig:tau_gg} shows the internal gamma-ray optical depth ($\tau_{\gamma\gamma}$) in the core region. The core region is expected to be optically thick against gamma-ray photons above 10--100~MeV depending on disk luminosities. Such high optical thicknesses against pair production in AGN coronae are well known \citep[e.g.,][]{1971MNRAS.152...21B, Done1989, Fabian2015} based on the compactness parameter argument \citep{Guilbert1983}.

For the gamma-ray attenuation in AGNs, we can consider two cases. One is the ``uniform'' emissivity case, while the other is the ``screened'' case. In the uniform emissivity case, gamma-rays and target photons are uniformly distributed. Gamma rays are attenuated by a factor of $3u(\tau_{\gamma\gamma})/\tau_{\gamma\gamma}$, where $u(\tau_{\gamma\gamma})=1/2 + \exp(-\tau_{\gamma\gamma})/\tau_{\gamma\gamma} - [1 - \exp(-\tau_{\gamma\gamma})]/\tau_{\gamma\gamma}^2$  \citep[See Sec. 7.8 in][]{Dermer2009}. In the screened case, gamma-rays are assumed to be generated in the inner part of the corona, and the dominant attenuating photon field surrounds it. Since the disk and corona temperature depends on the disk radius \citep{Kawanaka2008PASJ...60..399K}, such configuration can be realized. Then, gamma-rays are attenuated by a factor of $\exp(-\tau_{\gamma\gamma})$.

\begin{figure*}
 \begin{center}
  \includegraphics[width=0.7\linewidth]{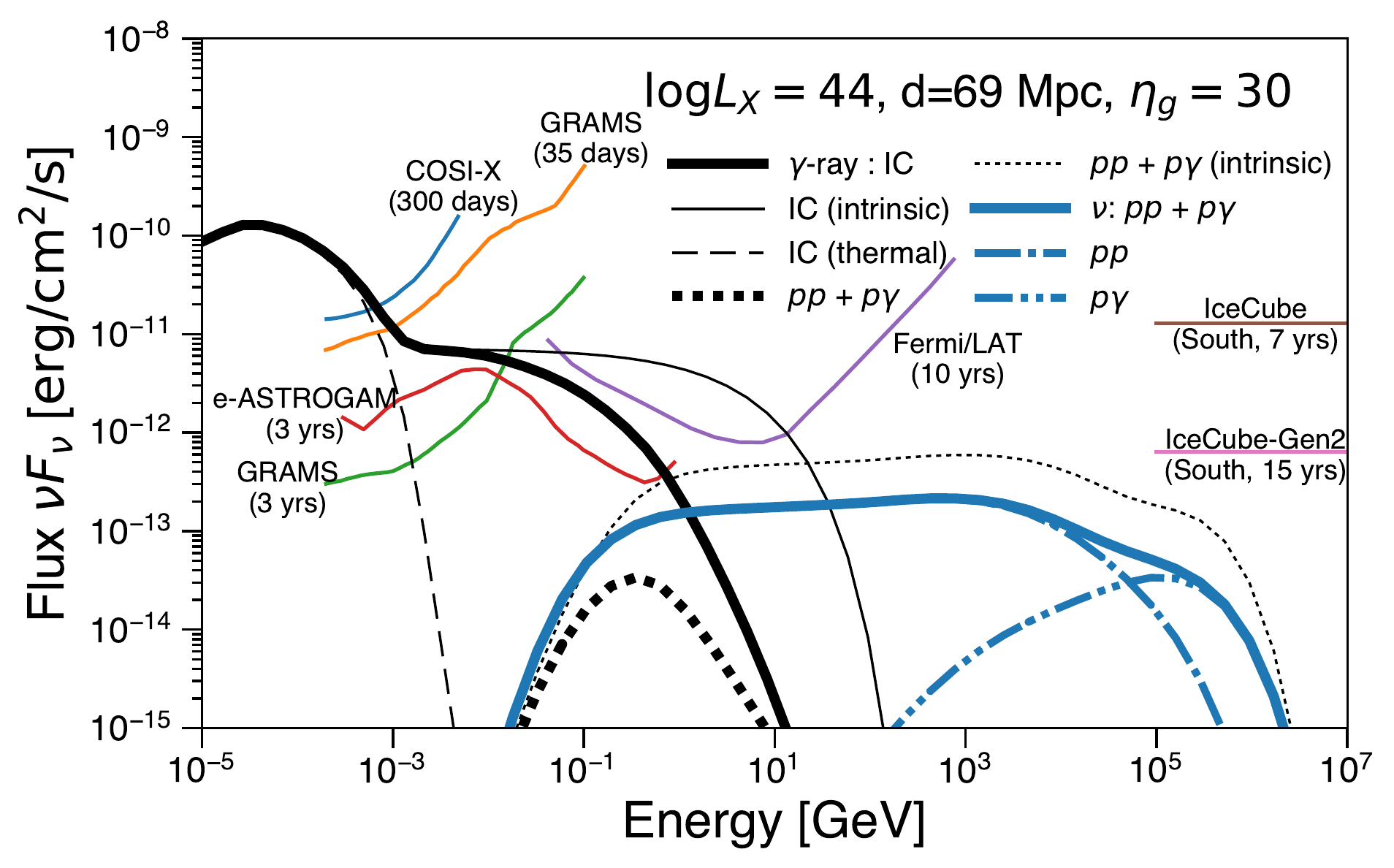}
\caption{Gamma-ray and neutrino spectrum per flavour from AGN coronae with $p_{\rm inj}=2.0$ and $\eta_g=30$. We set 2-10~keV luminosity of $10^{44}~{\rm erg\ s^{-1}}$ at a distance of 69~Mpc (IC~4329A). The thick black solid and thick dot curve shows gamma rays from IC interaction and $pp+p\gamma$ interaction including attenuation effect. Each thin curve shows the spectrum before the attenuation. Taken from \citet{Inoue2019ApJ...880...40I}.} \label{fig:SED_gamma_nu}
 \end{center}
\end{figure*}

\subsection{General SED Pictures}
Figure~\ref{fig:SED_gamma_nu} shows the resulting steady-state gamma-ray and neutrino spectra from AGN coronae for the case assuming IC~4329A ($L_X=10^{44}~{\rm erg\ s^{-1}}$ at a distance of 69~Mpc). The neutrino flux is shown in the form of per flavor. In this figure, the injection spectral index is $p_{\rm inj}=2.0$, and the gyrofactor is $\eta_g=30$ for both electrons and protons. We also set the same power injected in protons and electrons. For the gamma-ray attenuation, the uniform emissivity model is adopted. 

Since the spectral index of electrons becomes $\sim3$ after radiative cooling, the resulting non-thermal gamma-ray spectrum is flat in $\nu F_\nu$ in the MeV band after the thermal cutoff at $\gtrsim1$~MeV. Due to the internal gamma-ray attenuation effect, the spectra will have a cutoff of around 100~MeV.

Gamma rays and neutrinos induced by hadronic interactions carry $1/3$ and $1/6$ of those interacted hadron energies. Also, the $pp$ and $p\gamma$ production efficiency are given by the ratio between the dynamical timescale and the interaction timescales. The $pp$ production efficiency is analytically given as $f_{pp}={t_{\rm fall}}/{t_{pp}} \simeq 0.16 ({\tau_{\rm T}}/{1.1})({r_c}/{40})^{-0.5}$. Therefore, hadronic gamma-ray and neutrino luminosity is expected to be $\sim5$\% and $\sim3$\% of the intrinsic proton luminosity. Since we assume the same energy injection to electrons and protons, hadronic gamma-ray and neutrino fluxes are $\sim5$\% and $\sim3$\% of the IC gamma-ray flux. 

Contrary to gamma rays, neutrinos induced by hadronic interactions can escape from the system without attenuation. Since we adopt the same $p_{\rm inj}=2$ for protons as for electrons, we expect a flat spectrum for neutrinos, to which $pp$ makes the dominant contribution. The exact position of the cutoff energy depends on the assumed $\eta_g$. Here, as described later, we set $\eta_g=30$ in order to be consistent with the IceCube background flux measurements. This gyrofactor results in a neutrino spectral cutoff around 100~TeV.

\subsection{Application to NGC~1068}
A nearby Seyfert galaxy NGC~1068 was reported as the hottest neutrino spot with a 2.9-$\sigma$ confidence level in 10-yr all-sky survey observations of IceCube \citep{IceCube2020PhRvL.124e1103A}. As we describe above, the hybrid corona model is one possible solution. Thus, understanding the required physical parameters to explain the neutrino in NGC~1068 is crucial for the coronal model test.

\begin{figure}[t]
 \begin{center}
  \includegraphics[width=0.7\linewidth]{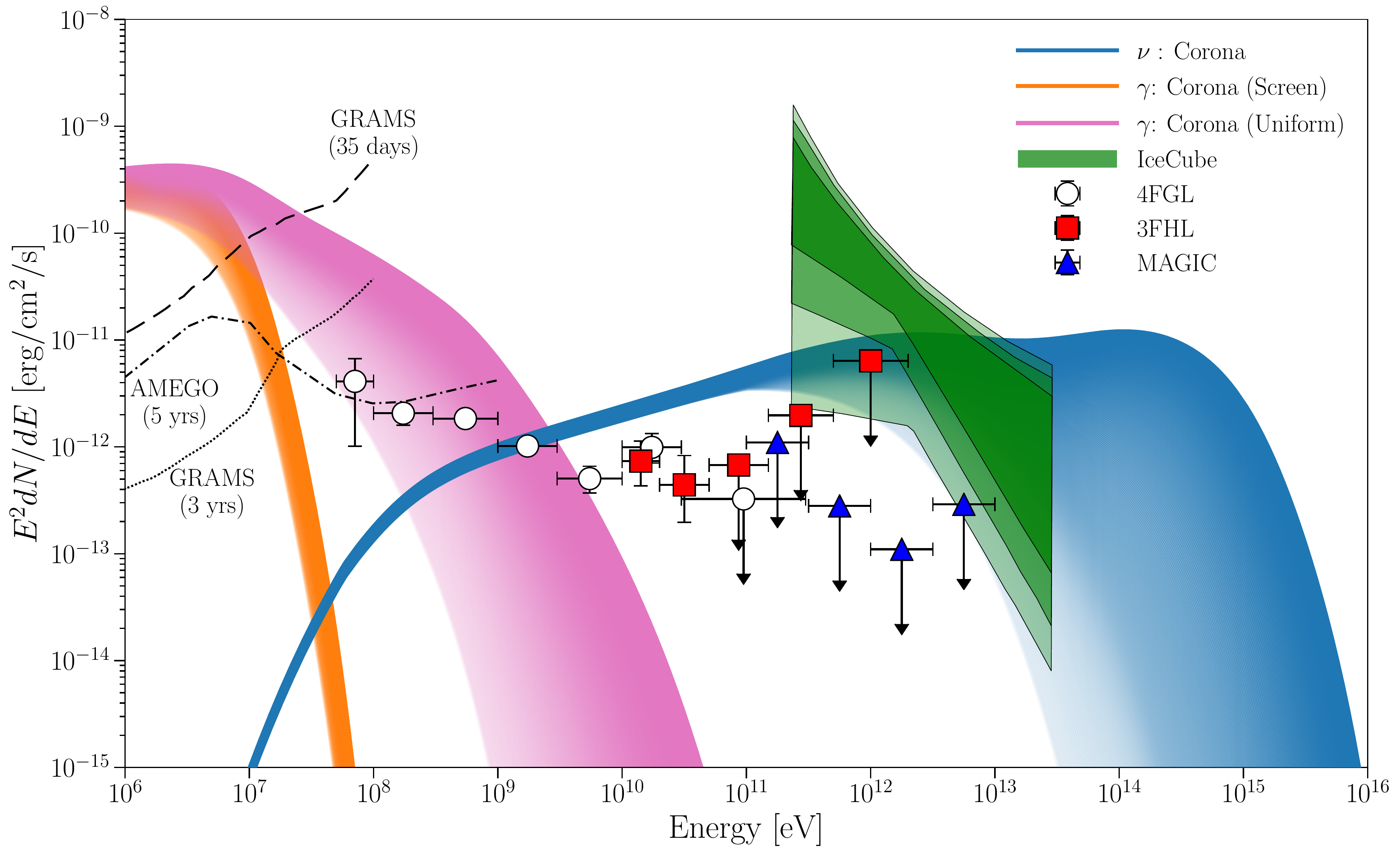}
\caption{The gamma-ray and neutrino spectrum of NGC~1068. The circle, square, and triangle data points are from \citet{4FGL}, \citet{3FHL}, and \citet{MAGIC2019ApJ...883..135A}, respectively. The green shaded regions represent the 1, 2, and $3\sigma$ regions on the spectrum measured by IceCube  \citep{IceCube2020PhRvL.124e1103A}. The expected gamma-ray and neutrino spectrum from the corona are shown for $30\le\eta_g\le3\times10^4$. The darker region corresponds to lower $\eta_g$. The blue region shows the expected neutrino spectrum. The orange and magenta shaded region shows the gamma-ray spectrum for the uniform case and the screened case, respectively. Taken from \citet{Inoue2020ApJ...891L..33I}.}\label{fig:GN_1}
 \end{center}
\end{figure}

Fig.~\ref{fig:GN_1} shows the expected gamma-ray and neutrino signals from NGC~1068 together with the observed gamma-ray data \citep{4FGL,3FHL,MAGIC2019ApJ...883..135A} and the IceCube data \citep{IceCube2020PhRvL.124e1103A}. We follow the assumptions on the coronal parameters as described above except for the gyrofactor and parameters determined by the coronal synchrotron model explaining the mm-excess (See \S.~\ref{sec:history} and Fig.~\ref{fig:ALMA}). Considering the neutrino measurement uncertainty, in the figure, we plot the model curve region in the range of $30\le\eta_g\le3\times10^4$ for each curve. The darker region corresponds to lower $\eta_g$, in which models extend to higher energies. Further detailed neutrino spectrum will narrow down the range of allowed $\eta_g$.

Regarding the gamma-ray measurements, in the screened gamma-ray attenuation case, the model can explain the preliminary neutrino signals above several~TeV without violating the gamma-ray data. On the other hand, the uniform emissivity model violates the low-energy gamma-ray data. This implies a further detailed study of coronal geometry is necessary. In either case, it is not easy for the corona model to explain the entire observed gamma-ray flux data up to 20~GeV, requiring another mechanism to explain gamma-rays above 100~MeV such as star formation activity \citep{Fermi2012ApJ...755..164A}, jet \citep{Lenain2010}, or disk wind \citep{Lamastra2016}. Therefore, the coronal model can explain the IceCube neutrino signal without violating the gamma-ray data.

An important question is what differentiates NGC~1068 from other nearby Seyfert galaxies. NGC~1068 is not the brightest X-ray Seyfert \citep{Oh2018}. Its observed hard X-ray flux is a factor of $\sim16$ fainter than the one of the brightest Seyfert, NGC~4151. NGC~1068 is a type-2 Seyfert galaxy, and obscured by the materials up to the neutral hydrogen column density of $N_H\sim10^{25} {\rm cm^{-2}}$ \citep{Bauer2015, Marinucci2016}. If we correct this attenuation effect to understand the intrinsic X-ray radiation power, NGC~1068 appears to be the intrinsically brightest Seyfert. For example,  intrinsically, it would be by a factor of $\sim3.6$ brighter than NGC~4151 in X-ray. As the dusty torus does not obscure coronal neutrino emission, which can scale with accretion power, NGC~1068 might be the brightest source in the neutrino sky. This could be why NGC~1068 appears as the hottest spot in the IceCube map rather than other Seyfert galaxies.
 
 In NGC~1068, the jets are prominent and extend for several kpcs in both directions. In the central $\sim14-70$~pc region, the downstream jet emission dominates in the centimeter regime \citep{Gallimore1996, Gallimore2004}. These jets can also be the production site of the reported neutrinos. However, gamma-ray attenuation is not significant in these far side regions from the nucleus. Therefore, these jets may not be the dominant neutrino production sites.
 
\section{Cosmic Gamma-ray and neutrino background radiation}
\label{sec:CGNB}

\begin{figure}
 \begin{center}
	\includegraphics[width=0.7\linewidth]{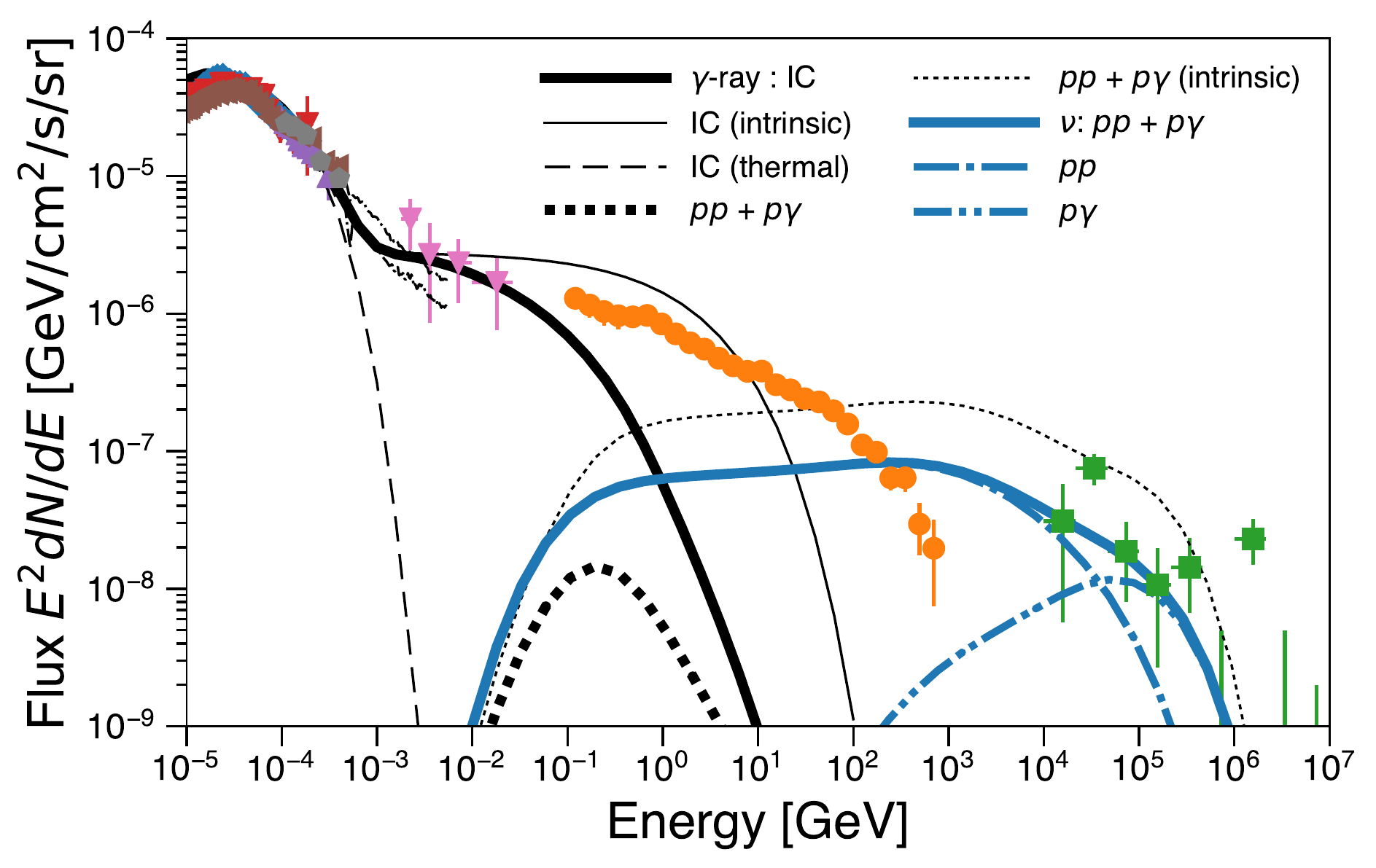}
\caption{The cosmic gamma-ray and neutrino background spectrum from AGN coronae with $p_{\rm inj}=2.0$ and $\eta_g=30$. Each component is labelled as in the figure. Thin curves show the spectra before the attenuation. The circle-orange and square-green data points correspond to the total cosmic gamma-ray background spectrum measured by the {\it Fermi}/LAT \citep{Fermi_CGB_2015ApJ...799...86A} and the cosmic neutrino background spectrum by the IceCube \citep{Aartsen2015}, respectively. The cosmic X-ray and MeV gamma-ray background spectrum data of  {\it HEAO}-1 A2 \citep[triangle-brown;][]{gru99}, {\it INTEGRAL} \citep[triangle-red;][]{chu07}, {\it HEAO}-1 A4 \citep[triangle-purple;][]{kin97}, \textit{Swift}-BAT \citep[diamond-blue][]{aje08}, {\it SMM} \citep[black-thin-dot-dashed][]{Watanabe1997AIPC..410.1223W}, Nagoya--Ballon \citep[pentagon-gray;][]{Fukada1975Natur.254..398F}, COMPTEL \citep[pink-triangle;][]{Weidenspointner2000AIPC..510..467W} are also shown in the figure.  Taken from \citet{Inoue2019ApJ...880...40I}.}\label{fig:CGNB_etag_100}
 \end{center}
\end{figure}

In this section, we consider the cosmic gamma-ray and neutrino background spectra from AGN coronae. For the details of calculation, the readers may refer to \citet{Inoue2019ApJ...880...40I}. Fig.~\ref{fig:CGNB_etag_100} shows the cosmic X-ray/gamma-ray and neutrino background spectra from AGN coronae assuming the case of $p_{\rm inj}=2.0$ and $\eta_g=30$ together with the observed background spectrum data by {\it HEAO}-1 A2 \citep{gru99}, {\it INTEGRAL} \citep{chu07}, {\it HEAO}-1 A4 \citep{kin97}, \textit{Swift}-BAT \citep{aje08}, {\it SMM} \citep{Watanabe1997AIPC..410.1223W}, Nagoya--Balloon \citep{Fukada1975Natur.254..398F}, COMPTEL \citep{Weidenspointner2000AIPC..510..467W}, {\it Fermi}-LAT \citep{Fermi_CGB_2015ApJ...799...86A}, and IceCube \citep{Aartsen2015}. 

By setting $f_{\rm nth}=0.03$, AGN coronae can nicely explain the cosmic MeV gamma-ray background in an extension from the cosmic X-ray background radiation. Since the spectral index of non-thermal electrons in the coronae is $\sim3$, the resulting MeV gamma-ray background spectrum becomes flat in $E^2dN/dE$. Since the dominant IC contributors switch from thermal electrons to non-thermal electrons at around $1$~MeV, the MeV background spectrum may have a spectral hardening feature at $\sim1$~MeV. 

In Fig.~\ref{fig:CGNB_etag_100}, we set $\eta_g=30$. The result for the MeV background does not significantly change as far as $\eta_g<1000$. If $\eta_g>1000$, we may require lower $f_{\rm nth}$. Here, IC emission due to non-thermal electrons also contributes to the X-ray band. Their contribution is about $\sim5$\% at 30~keV of the observed cosmic X-ray background flux, which may reduce the required number of the Compton-thick population of AGNs. Due to the internal gamma-ray attenuation effect, these non-thermal gamma rays can not contribute to the emission above GeV, where blazars, radio galaxies, and star-forming galaxies dominate \citep[see, e.g.,][]{Ajello2015}.

For neutrinos, the combination of $pp$ and $p\gamma$ interactions can nicely reproduce the IceCube fluxes below 100--300~TeV.  $pp$ interactions dominate the flux at $\lesssim10$~TeV, while $p\gamma$ interactions prevail above this energy. Because of the target photon field SED, $p\gamma$ is subdominant in the GeV-TeV band. If we inject more powers into protons, it inevitably overproduces the IceCube background fluxes.

\begin{figure}
 \begin{center}
  \includegraphics[width=0.7\linewidth]{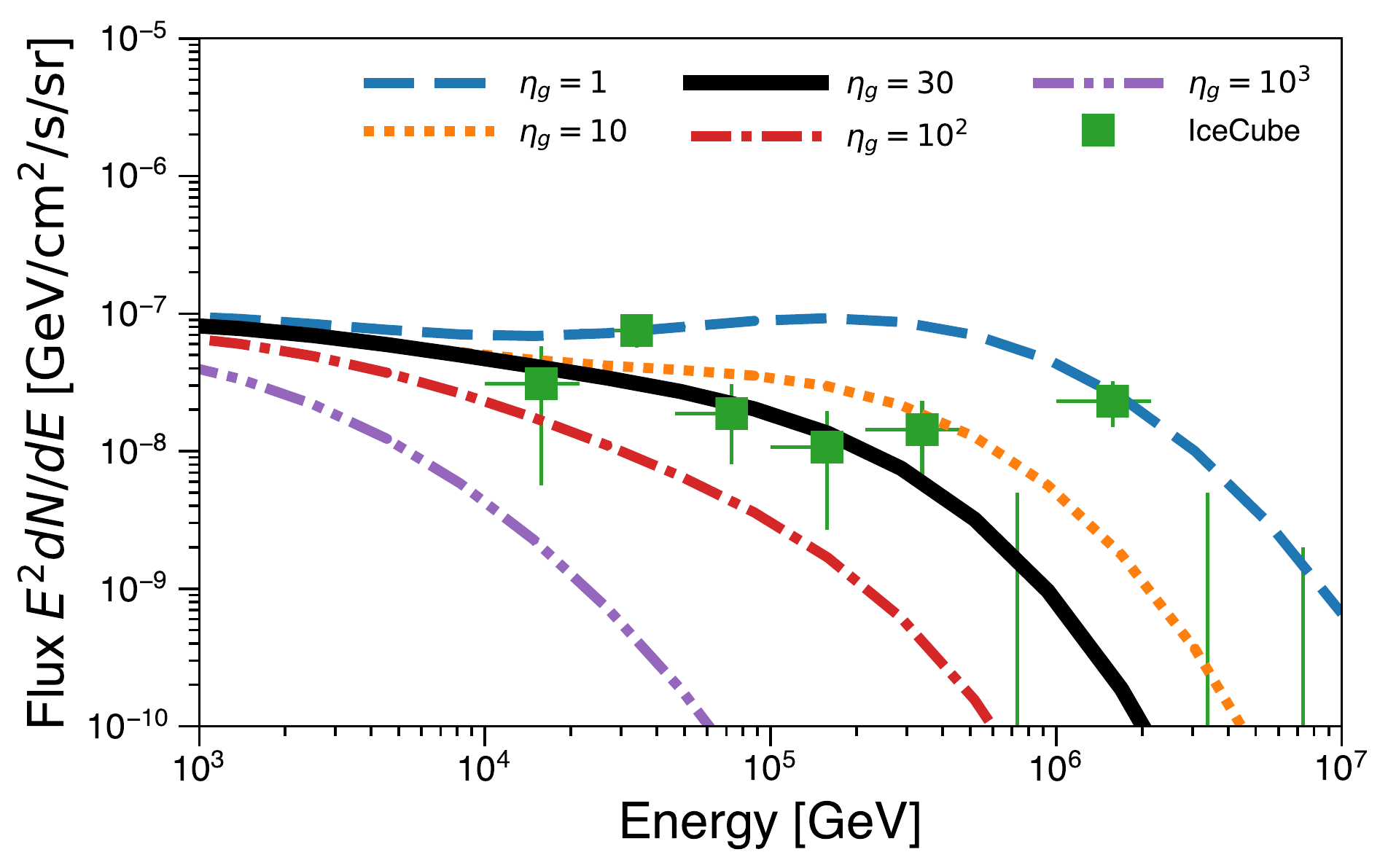}
\caption{The cosmic neutrino background spectrum per flavour from AGN coronae. The dashed, dotted, solid, dot-dashed, and double-dot-dashed curve shows the contribution with $\eta_g=$1 (Bohm limit), 10, 30, $10^2$, and $10^3$, respectively. The square data points correspond to the cosmic neutrino background spectrum by the IceCube \citep{Aartsen2015}. Taken from \citet{Inoue2019ApJ...880...40I}.}\label{fig:CNB_All}
 \end{center}
\end{figure}

Figure~\ref{fig:CNB_All} shows the cosmic neutrino background spectra from AGN cores with various gyro factors ranging from 1 (Bohm limit) to $10^3$. It is clear that if $\eta_g\ll30$, the resulting neutrino fluxes overproduce the measured fluxes. On the contrary, if $\eta_g\gg30$, AGN coronae can not significantly contribute to the observed neutrino background fluxes. Thus, to explain the IceCube neutrino background fluxes by AGN cores, $\eta_g\sim30$ is required.

\section{Discussion}
\label{sec:discussion}

\subsection{Role of Secondary Particles}
Secondary particles can be injected in corona by hadronic processes produce and by \(\gamma\gamma\) pair creation.  Even if these secondary particles are negligible by the number density, they still can be energetically important. Hadronic gamma-ray fluxes before the pair creation are about a factor of 10 less than that by primary electrons (see Fig.~\ref{fig:SED_gamma_nu}) because of radiative efficiency differences between protons and electrons. Therefore, hadronically induced secondary pairs should not be energetically important.

Pairs generated through $\gamma\gamma$ annihilation of primary leptonic fluxes may also contribute such as to the MeV gamma-ray spectrum. However, considering the intrinsic photon index $\Gamma_{\rm ph}=2$ and acceleration limit, the resulting pair contribution will not be higher than primaries. Thus, the secondary particle component would not significantly alter the resulting spectra. 

Here, hadronic power can be much higher than we assume. Then, secondary leptons can energetically dominate the primary leptons. However, as seen in Fig.~\ref{fig:CGNB_etag_100}, such models can not explain the cosmic MeV gamma-ray and TeV-PeV neutrino background fluxes.

\subsection{Comparison of current available models}
\label{sec:comp}

\begin{figure}
 \begin{center}
  \includegraphics[width=0.7\linewidth]{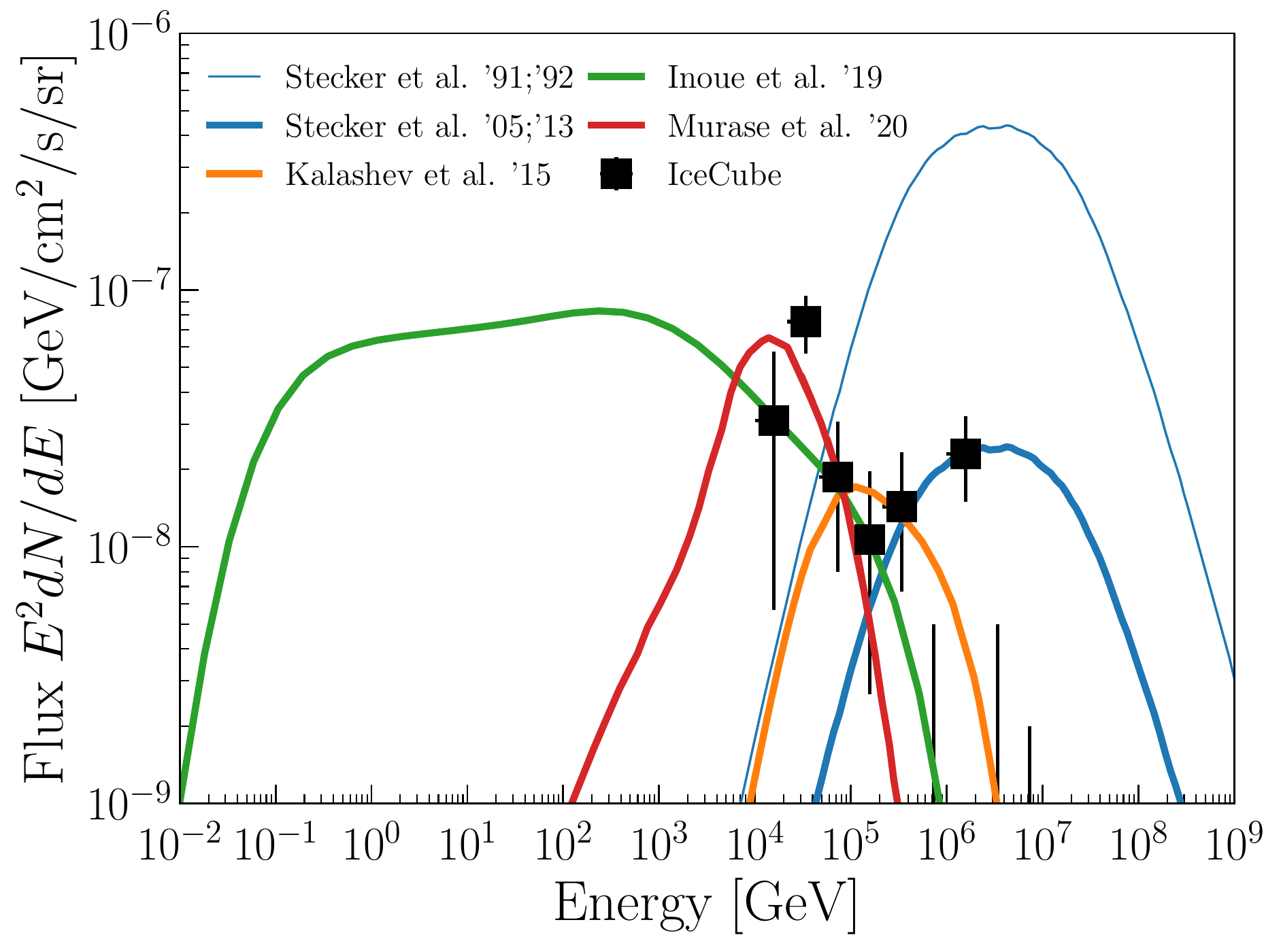}
\caption{The cosmic neutrino background spectrum per flavour from AGN coronae, showing the currently available models \citep{Stecker2013,Kalashev2015,Inoue2019ApJ...880...40I,Murase2020PhRvL.125a1101M}. The original prediction by \citet{Stecker1991PhRvL..66.2697S, Stecker1992} is also shown.}\label{fig:CNB_comp}
 \end{center}
\end{figure}

In literature, it has been argued that high energy particles in the AGN coronae generate intense neutrino emission \citep[e.g.,][]{Eichler1979, Begelman1990, Stecker1992}. These originally predicted fluxes have been ruled out by the IceCube observations \citep{IceCube2005}. However, recent studies have revisited the estimated fluxes \citep{Stecker2013, Kalashev2015, Inoue2019ApJ...880...40I, Murase2020PhRvL.125a1101M}. In this section, we clarify the differences in recent AGN corona models for high-energy neutrinos. Figure~\ref{fig:CNB_comp} shows the resulting cosmic neutrino background spectrum from those papers. These corona models predict different neutrino spectra, although they consider the same neutrino production region. This section briefly describes the differences among those models.

\citet{Stecker2013} considered a similar model of the originally proposed one \citep{Stecker1992}, but the background flux is assumed to be lower by a factor of 20 to match with the IceCube flux. The shock radius and the magnetic field strength were assumed to be $10R_s$ and $10^3$~G in the model by \citet{Stecker1992}. The particle spectral index was also assumed to be 2 in the framework of the DSA.

\citet{Kalashev2015} followed the treatment in \citet{Stecker1992, Stecker2013}, but accounting for the radial emission profile in the standard accretion disk in their consideration of the $p\gamma$ cooling processes. The particle spectra in \citet{Kalashev2015} are normalized to match with the IceCube data. Given the observationally determined corona size $R_c\sim40R_s$, the dominant photon target is likely to be generated in the inner region of the coronae.

\citet{Inoue2019ApJ...880...40I} took into account both X-ray and radio measurements, which allowed us to derive $R_c=40R_s$ and $B_c=10$~G \citep{Inoue2018}. And, to explain the cosmic MeV background by Seyferts, $f_{\rm nth}$ is set as $0.03$. The proton spectral index is assumed to be the same as that of electrons, determined through the radio measurement. However, the non-thermal electron/proton energy ratio is set to be $K_{ep}=1$, and $\eta_g=30$ are assumed to reproduce the cosmic neutrino background fluxes. We note that it can be $30\le\eta_g\le3\times10^4$ for the explanation of NGC~1068. $R_c$, $B_c$, $p$, and $f_{\rm nth}$ are, for the first time, observationally determined, while $K_{ep}$ and $\eta_g$ are still assumed to reproduce the IceCube measurements. Besides, the exact acceleration process is not well determined, as they follow the radio observation results.

\citet{Murase2020PhRvL.125a1101M} consider stochastic acceleration motivated by recent numerical simulations for low-luminosity AGNs \citep{Kimura2019}. The required cosmic ray pressure to explain the IceCube data is about 1-10\% of thermal pressure, which is similar to \citet{Inoue2019ApJ...880...40I}. Their model assumed $R_c=30R_s$ and $B_c\simeq10^3$~G. This high magnetic field contradicts the radio measurements. As discussed above, the measured magnetic field may not be high enough to accelerate particles in the stochastic acceleration scheme efficiently. 

Very recently, \citet{Gutierrez2021arXiv210211921G} also model high energy signals from AGN coronae taking into account X-ray and radio observations. They basically follow \citet{Inoue2019ApJ...880...40I}. However, they consider lower $K_{ep}$ than in \citet{Inoue2019ApJ...880...40I}. This assumption is more consistent with $K_{ep}$ seen in nearby supernova remnants. Because of this low $K_{ep}$, secondary leptons make the dominant contribution in the MeV band. Although they have not estimated the integrated MeV gamma-ray and TeV-PeV neutrino background fluxes in their paper yet, such a study based on their model will be helpful for future comparison.

\subsection{Future test of models}

Although non-thermal AGN coronal models have failed to explain the X-ray data in the 1990s, millimeter radio observations found weak coronal activity in nearby Seyferts in 2018, and then the possible detection of NGC~1068 in 2020 shed light again on the AGN corona model. However, as described above, the models still have profound ambiguity. We require multi-messenger tests on that. Below we list possible ways to test the AGN corona scenario.

\subsubsection{Radio Synchrotron Emission}
As non-thermal particles are accelerated in the magnetized coronae, we should expect coronal synchrotron emission. ALMA observations have already grasped the evidence of the coronal synchrotron emission. Therefore, we need to consider the expected radio synchrotron emission for each model and compare it with the ALMA measurements. As the ALMA data are already available for nearby Seyferts, this test can be easily performed.

\subsubsection{Nuclear Spallation Effect Appearing in X-ray}
High energy protons can be traced by future high-resolution calorimeter spectroscopy in the X-ray band such as {\it XRISM} \citep{Tashiro2018} and {\it Athena} \citep{Nandra2013}. As narrow line features are seen in AGN X-ray disk spectra, there are abundant metal elements in AGN cores. Accelerated protons interact with those nuclei and induce nuclear spallation. The nuclear spallation in AGN disks will enhance emission lines from Mn, Cr, V, and Ti \citep{Gallo2019}. Those signatures will be another clue for the test of the corona model.

\subsubsection{MeV power-law tail}
In the hybrid AGN corona model, the non-thermal gamma-ray should appear after the thermal cutoff ($\sim300$~keV). The expected MeV gamma-ray flux is about 5\% of the intrinsic X-ray flux. Due to the strong internal gamma-ray attenuation effect, the spectra will have a cutoff around 10-100~MeV depending on the photon distribution. These MeV gamma-ray photons should be seen by future MeV gamma-ray observatories. If not, this means that the amount of non-thermal population is much less than required for the MeV gamma-ray background radiation, which means the corona model can not explain the neutrino background.

\subsubsection{Further Neutrino Observations}
The neutrino measurement can provide a critical test on AGN corona scenarios and robustly constrain acceleration of protons. Currently, only NGC~1068 is reported as the possible neutrino production site. Further neutrino observations should see more nearby Seyferts following the AGN corona model \citep{Inoue2019ApJ...880...40I, Kheirandish2021arXiv210204475K,Gutierrez2021arXiv210211921G}. Additionally, neutrino observations may pin down the acceleration mechanism in the AGN corona by taking the neutrino spectrum from individual objects.

\section{Summary}
\label{sec:summary}
In this {\it Review}, we overview our current understanding of the AGN coronae from X-ray and mm radio observations. We show that these measurements contain critical information for constraining particle acceleration in the AGN coronae. AGN coronae are feasible sites for particle acceleration. If the energy injection rate is adequate, AGN coronae can explain the neutrino signals from NGC~1068 and a part of the diffuse neutrino fluxes. Future radio, X-ray, MeV gamma-ray, and TeV neutrino observations will be able to test this scenario by observations of nearby bright Seyferts.
\vspace{6pt} 

\acknowledgments{YI is supported by JSPS KAKENHI Grant Number JP16K13813, JP18H05458, JP19K14772, program of Leading Initiative for Excellent Young Researchers, MEXT, Japan, and RIKEN iTHEMS Program. DK is supported by JSPS KAKENHI Grant Numbers JP18H03722, JP18H05463, and JP20H00153.}

\end{paracol}
\reftitle{References}


\externalbibliography{yes}
\bibliographystyle{mdpi}

%


\end{document}